\documentclass[12pt, draftclsnofoot, onecolumn]{IEEEtran}
\usepackage{multicol}  
\IEEEoverridecommandlockouts
\usepackage[pdftex]{graphicx}
\usepackage[ruled]{algorithm2e}
\usepackage{algorithmic}
\usepackage[small]{caption}
\usepackage{float}
\usepackage{amsfonts}
\usepackage{xspace}
\usepackage{fancyhdr}
\usepackage{amssymb,amsmath}
\usepackage{subfig}
\usepackage{epstopdf}
\usepackage{bbding}
\usepackage{cite} 
\usepackage{color}
\begin{document}

\title{Rate Splitting Multiple Access for Next Generation Cognitive Radio Enabled LEO Satellite Networks}
\author{Wali Ullah Khan, \textit{Member, IEEE,} Zain Ali, Eva Lagunas, \textit{Senior Member, IEEE,} Asad Mahmood, Muhammad Asif, Asim Ihsan, Symeon Chatzinotas, \textit{Fellow, IEEE,} Bj\"orn Ottersten, \textit{Fellow, IEEE,} and Octavia A. Dobre, \textit{Fellow, IEEE}\thanks{This work has been supported by the Luxembourg National Research Fund (FNR) under the project MegaLEO (C20/IS/14767486).  An earlier version of this paper was presented at the main symposium of IEEE GLOBECOM 2022, Brazil \cite{Conf}. 

Wali Ullah Khan, Eva Lagunas, Symeon Chatzinotas, and Bj\"orn Ottersten are with the Interdisciplinary Center for Security, Reliability and Trust (SnT), University of Luxembourg, 1855 Luxembourg City, Luxembourg (e-mails: \{waliullah.khan, eva.lagunas, asad.mahmood, symeon.chatzinotas, bjorn.ottersten\}@uni.lu).

Zain Ali is with Department of Electrical and Computer Engineering, University of California, Santa Cruz, USA (e-mail: zainalihanan1@gmail.com).

Muhammad Asif is with Guangdong Key Laboratory of Intelligent Information Processing, College of Electronics and Information Engineering, Shenzhen Univerisyt, Shenzhen, Guangdong China (e-mail: masif@szu.edu.cn).

Asim Ihsan is with the School of Computer Science and Electronic Engineering, Bangor University, Bangor LL57 1UT, U.K. (email: a.ihsan@bangor.ac.uk).

Octavia A. Dobre is with the Department of Electrical and Computer Engineering, Faculty of Engineering and Applied Science,  Memorial University, St. John's, NL A1B 3X9, Canada (e-mail: odobre@mun.ca).  

}}%

\maketitle

\begin{abstract}
Low Earth Orbit (LEO) satellite communication (SatCom) has drawn particular attention recently due to its high data rate services and low round-trip latency. It has low launching and manufacturing costs than Medium Earth Orbit (MEO) and Geostationary Earth Orbit (GEO) satellites. Moreover, LEO SatCom has the potential to provide global coverage with a high-speed data rate and low transmission latency. However, the spectrum scarcity might be one of the challenges in the growth of LEO satellites, impacting severe restrictions on developing ground-space integrated networks. To address this issue, cognitive radio and rate splitting multiple access (RSMA) are the two emerging technologies for high spectral efficiency and massive connectivity. This paper proposes a cognitive radio enabled LEO SatCom using RSMA radio access technique with the coexistence of GEO SatCom network. In particular, this work aims to maximize the sum rate of LEO SatCom by simultaneously optimizing the power budget over different beams, RSMA power allocation for users over each beam, and subcarrier user assignment while restricting the interference temperature to GEO SatCom. The problem of sum rate maximization is formulated as non-convex, where the global optimal solution is challenging to obtain. Thus, an efficient solution can be obtained in three steps: first we employ a successive convex approximation technique to reduce the complexity and make the problem more tractable. Second, for any given resource block user assignment, we adopt Karush–Kuhn–Tucker (KKT) conditions to calculate the transmit power over different beams and RSMA power allocation of users over each beam. Third, using the allocated power, we design an efficient algorithm based on the greedy approach for resource block user assignment. For comparison, we propose two suboptimal schemes with fixed power allocation over different beams and random resource block user assignment as the benchmark. Numerical results provided in this work are obtained based on the Monte Carlo simulations, which demonstrate the benefits of the proposed optimization scheme compared to the benchmark schemes.   
\end{abstract}


\section{Introduction} 
Satellite communication (SatCom) has recently gained significant attention in industry and academia due to its capability to provide global coverage and support a wide range of services \cite{9460776}. Three existing SatCom types are the Geostationary Earth Orbit (GEO) satellite, the Medium Earth Orbit (MEO) satellite, and Low Earth Orbit (LEO) satellite. Due to the low orbit profile, the LEO SatComhas the ability to provide high-speed data and connect massive ground users with low round-trip latency \cite{9512414}. Moreover, its manufacturing and launching costs are comparatively lower than GEO and MEO satellite systems, making it more likely to achieve global coverage. However, the increasing demand for different services would require many LEO satellites in orbit, which can be challenging using limited spectrum resources \cite{9222141}. In such a spectrum scarcity situation, satellites in different orbits would need to reuse the same spectrum and cover the same geographical area on the earth, creating the issue of mutual interference. Therefore, allocating spectrum efficiently among satellites in different orbits can be a critical issue if not adequately resolved. As a result, the future development of large-scale LEO SatCom networks can be seriously affected. In this regard, researchers are actively pursuing and exploring new approaches for developing next-generation LEO SatCom \cite{7060478}. One of the potential approaches which can be helpful in this situation is efficient spectrum sharing across different orbits using advanced spectrum allocation techniques \cite{9210567}.
%%%%%%%%%%%%%%%%%%%%%%%%%%%%%%%%%%%%%%%%%%%%%%%%%
\begin{table}[h]
\centering
\caption{Different symbols and definition}
\begin{tabular}{|c|c|} \hline 
\textbf{Symbol} & \textbf{Definition}\\
\hline
$K$ & Number of resource blocks \\ \hline
$M$ & Number of LEO satellite beams  \\ \hline
$U$ & Set of total LEO users \\ \hline
$U_m$ & Set of users over $m$ beam \\ \hline
$k$ & Index of $k$ resource block \\ \hline
$m$ & Index of $m$ beam \\ \hline
$u$ & Index of $u$ user in set $N$ \\ \hline
$s_{m,k}$ & Overall signal over $k$ resource block \\ \hline
$s_{m,0,k}$ & Common data symbol over $k$ resource block \\ \hline
$s_{m,u,k}$ & Private data symbol over $k$ resource block \\ \hline
$p_{m,k}$ & Transmitted power of $m$ beam over $k$ \\ \hline
$\eta_{m,0,k},\eta_{m,u,k}$ & RSMA power allocation coefficients \\\hline
$h_{m,u,k}$ & Channel gain from $m$ beam to $u$ user\\ \hline 
$G_T$ & Gain of transmit antenna \\ \hline 
$G_R$ & Gain of received antenna \\ \hline 
$D$ & Distance from satellite to user \\ \hline 
$c$ & Speed of light \\ \hline 
$f_c$ & Carrier frequency \\ \hline 
$y_{m,u,k}$ & Received signal at user \\ \hline 
$g_{u,k}$ & Channel gain from GEO to $u$ user over $k$\\ \hline
$q_{u,k}$ & Transmit power of GEO over $k$ \\ \hline  
$e_{j,k}$ & Signal of $j$ GEO user over $k$ \\ \hline 
$\omega_{m,u,k}$ & Additive white Gaussian noise. \\ \hline 
$W$ & Bandwidth available at LEO satellite \\ \hline
$R_{m,c,k}$ & Data rate of common signal \\ \hline
$x_{m,u,k}$ & Assignment variable of user \& resource block \\ \hline   
$R_{m,u,k}$ & Data rate of private signal \\ \hline 
$\sigma^2$ & Noise variance \\ \hline 
$I_{R_{m,c,k}}$ & RSMA user interference for common signal \\ \hline 
$I_{R_{m,u,k}}$ & RSMA user interference for private signal \\ \hline 
$f_{m,j,k}$ & Channel gain from LEO to GEO user \\ \hline 
$I_{th}$ & Threshold of maximum interference temperature \\ \hline 
$R_{min}$ & Threshold of minimum data rate \\ \hline 
$P_{tot}$ & Total power budget of LEO satellite\\ \hline
$\delta$ & Positive step size \\ \hline 
\end{tabular}
\label{tab1}
\end{table}
%%%%%%%%%%%%%%%%%%%%%%%%%%%%%%%%%%%%%%%%%%%%%%%%%

Cognitive radio and rate splitting multiple access (RSMA) have emerged as the promising technologies for providing high spectral efficiency and have the potential to ease the above situation \cite{8957541,8357810}. These technologies can simultaneously accommodate multiple users over the same spectrum and time resources, which significantly enhances the system connectivity. More specifically, in cognitive radio, the licensed primary and unlicensed secondary networks communicate over the same spectrum such that the secondary network would not cause harmful interference to the primary network \cite{6845054}. On the other side, the fundamental concept of RSMA is to accommodate multiple users over the same spectrum and time resources. According to RSMA protocol, the signal transmitted to the users is divided into two signals, i.e., the common part and the private part \cite{9461768}. The common parts of the signals can be combined into a single common signal first and then it encoded with a public shared code-book. In contrast, the private parts of the signals can be independently encoded to specific intended users. Each user first decodes the common part of the signal using the shared code-book. Then users reconstruct the original signals from the part of their common and intended private signals using the successive interference cancellation (SIC) technique.

The most common schemes of RSMA are 1-layer RSMA, 2-layer RSMA and generalized RSMA, respectively. These RSMA schemes use linear precoding techniques. 1-layer rate splitting, the foundation of nearly all RSMA schemes, is the simplest and most practical RSMA scheme currently in use. Extensive research has been conducted on 1-layer rate splitting. The transmitter activated the message combiner and used linear precoding in accordance with its principles. Every message is divided into a public and a private version. All users' shared messages are merged into a single one and encoded into a single stream using a universal codebook. However, each user's private communications are encoded in their own separate private stream. And while everyone can decode the public message, only the intended recipients can read the private ones. The 2-layer RSMA scheme was initially proposed for frequency division duplex massive multiple input multiple output to improve the system's robustness and achievable rate. On the basis of its principle, the system categorizes its users into different sets. A user's message was divided into an outer group common message, an inner group message, and a private message in each group. Then, everyone in the network receives the same inter-group message that has been encoded into a standard data feed using the same codebook. When members of a group exchange messages with one another, they combine those messages into a single message that is then encoded into a common stream for the group using a secret codebook. Only people who use it can crack the code. Private messages are encrypted in several separate private streams, each of which can only be read by its corresponding user.

%%%%%%%%%%%%%%%%%%%%%%%%%%%%%%%%%%%%%%%%%%%%%%%%%%%%%%%%%%%%%%%%%%%%%%%%%%%%%%%%%%%%%%%%%%%%%%%%%%%%%%%%%%%%
\subsection{Recent Works}
The spectrum sharing methods in non-terrestrial networks has gained much interest due to the limited resources. For instance, the authors of  \cite{8641187,5581292} have investigated co-channel interference in the coexistence network of LEO and GEO for both uplink and downlink communication, as well as investigated the effect of isolation angle on interference magnitude. Following that, to mitigate the interference between LEO and GEO SatCom networks and efficiently utilize of shared spectrum, the works in \cite{sharma2016line,8928079,9347950} have applied an adaptive power control technique to maximize the throughput of the system. Their proposed power control technique substantially impacts the throughput of the LEO SatCom network.
In a similar fashion, the authors in \cite{8352660} have used beam hopping and power control to realize LEO and GE spectrum sharing. They maximize the spectral efficiency of GEO SatCom while restricting the interference temperature to LEO SatCom. Moreover, Zhang {\em et al.} \cite{8398221} have improved the performance of phased array antenna through a nonlinear programming problem such that the quality of services of LEO SatCom and interference to GEO SatCom are protected. The works in \cite{Jalali,wang2020co} have also analysed interference among GEO and non-GEO SatCom systems. Further, Reference \cite{9512414} has solved a continuous power control problem for dynamic spectrum sharing and cooperative services between GEO and LEO SatCom networks.

%In SatCom, resource allocation and user grouping are critical issues. However, as seen in the preceding article, limited resources like as transmit power and bandwidth are critical characteristics that bottleneck the performance of the LEO satellites' communication system. Furthermore, clustering a large number of users raises computing complexity significantly.
Besides the above works that focus on spectral coexistence in non-terrestrial networks, some researchers have proposed non-orthogonal multiple access (NOMA) as a promising alternative for providing higher spectral efficiency and massive connectivity than classic orthogonal multiple access (OMA) approaches. NOMA allows multiple users to communicate over the same spectrum simultaneously. For this, the authors in \cite{9419053} have considered NOMA in the multiple input multiple output (MIMO) SatCom system to design and optimize the precoding vector and transmit power. Similarly, the work in \cite{8473479} has addressed the principles, methodology, and challenges of NOMA-based multi-satellite relay transmission. Moreover, the authors of \cite{7972935} have proposed an optimization algorithm for the user grouping technique and the transmit power allocation for the SatCom network using the NOMA technique. Following that, it is perceived that the sum rate performance in \cite{8642812} is better than \cite{7972935}, which uses a fixed beamforming method. Further, to reduce the completion time, the authors in \cite{9312147} have used the concept of cooperative NOMA in LEO SatCom networks. To reduce the power consumption of LEO SatCom, Yin {et al.} \cite{9165811} have employed NOMA with defective SIC for robust beamforming.

Recently, it has been proved that RSMA outperforms NOMA in providing high spectral efficiency \cite{9831440}. Although RSMA is extensively integrated to the terrestrial networks, some researchers have also considered RSMA for SatCom networks. For example, authors in \cite{9145200,9257433} have considered RSMA for GEO satellite communication and solved the max-min fairness problem using Weighted Minimum-Mean Square Error approach. Similarly in \cite{9324793}, the authors have considered RSMA for GEO unmanned aerial vehicle integrated networks. They investigate the sum rate maximization problem using sequential convex approximation and the first-order Taylor expansion approaches. Moreover, the works in \cite{9473795,9684855} have investigated the max-min data rate problems in GEO SatCom networks using RSMA technique. They consider RSMA beamforming and two-stage precoding schemes with imperfect channel state information. 

\subsection{Motivation and Contributions}
Based on the detailed literature, it can be observed that \cite{8641187,5581292,sharma2016line,8928079,9347950,8352660,8398221,9512414} provide spectrum sharing method, however they do not consider RSMA. Moreover, researchers of \cite{9419053,8473479,7972935,8642812,9312147,9165811} employ NOMA as a multiple access technique instead of RSMA technique. Further, the works in \cite{9145200,9257433,9324793,9473795,9684855} consider RSMA only in GEO satellite communication and they do not consider cognitive radio. Based on the available literature, the work that considers cognitive radio approach for spectral coexistence of GEO and LEO SatCom network using RSMA technique has not been investigated yet. To fill this potential research gap, this work consider cognitive radio enabled LEO SatCom using RSMA. The objective of this work is to maximize the spectral efficiency of the system under different practical constraints. In particularly, we simultaneously optimize the power budgets over the beams of LEO satellite, power allocation coefficients for ground users over each beam based on RSMA protocol, and resource block user assignment. The optimization framework is subjected to the interference temperature at GEO SatCom from LEO SatCom and the quality of services of LEO users. The formulated problem of spectral efficiency is not convex and obtaining a global optimal solution is challenging. Thus, we first adopt the successive convex approximation technique to make the problem more tractable, where a properly chosen surrogate can efficiently replace the original non-convex function. Then we apply Karush–Kuhn–Tucker (KKT) conditions for power allocation and an efficient algorithm for subcarrier beam assignment based on the greedy approach. We also propose two suboptimal schemes with a fixed power budget at each beam and random subcarrier beam assignment as the benchmark. The main contributions of our paper can be summarized as follow.
\begin{itemize}
    \item We proposes a cognitive radio enabled for LEO SatCom network using RSMA technique. Specifically, we consider that LEO SatCom operates as a secondary network with the existing primary GEO SatCom by sharing the same spectrum resources. To ensure the quality of services of primary GEO SatCom network, the proposed optimization framework restricts the interference temperature from secondary LEO SatCom to GEO SatCom network. The primary GEO SatCom uses orthogonal multiple access (OMA) technique to communicate with ground users using multiple resource blocks. The same resource blocks are then solely reused by the beams of LEO SatCom such that each beam uses one resource block to accommodate ground users using RSMA protocol. This work aims to maximize the sum rate of LEO SatCom network by optimizing various network resources such as power budget of different beams at LEO satellite, RSMA power allocation users over each beam, and resource block user assignment. 
    \item The optimization problem of sum rate maximization guarantees the quality of services of LEO users while restricts the interference temperature to GEO users. The optimization problem is formulated as non-convex and coupled with multiple variables, which makes it very challenging to obtain the global optimal solution. Therefore, we obtain a suboptimal yet efficient solution in three steps: first we adopt successive convex approximation technique to reduce the complexity and make the problem tractable. Second, for any given resource block user assignment, we exploit KKT conditions to obtain the closed-form solution over each beam. Third, using the calculated value of power over different beams and RSMA users, we designed an efficient algorithm for resource block user assignment using the greedy approach.
    \item To evaluate the performance of the proposed framework, numerical results are also provided and discussed. For fair comparison, we also provided two benchmark optimization frameworks with fixed power allocation over different beams and random resource block user assignment. In particular, the first benchmark optimizes transmit power of RSMA users and resource block user assignment while distributing the power of satellite over different beams equally. The second benchmark optimizes the transmit power over different beams and RSMA power allocation while assigning resource block user randomly. Results demonstrate the benefits of the proposed optimization framework for cognitive radio enabled LEO SatCom network compared to the other two benchmark frameworks. In addition, our proposed technique is low complex and converges after reasonable number of iterations. 
\end{itemize}
The reminder of this work can be structured as follow. Section II discusses channel and system model of cognitive radio enabled LEO SatCom network, followed by sum rate maximization problem. Section III provides the proposed optimization solution and Section IV provides and discusses the numerical results. Finally, Section V end up this paper with concluding remarks. Different notations used in this work are listed in Table 1.

%%%%%%%%%%%%%%%%
\begin{figure*}[!t]
\centering
\includegraphics[width=0.65\textwidth]{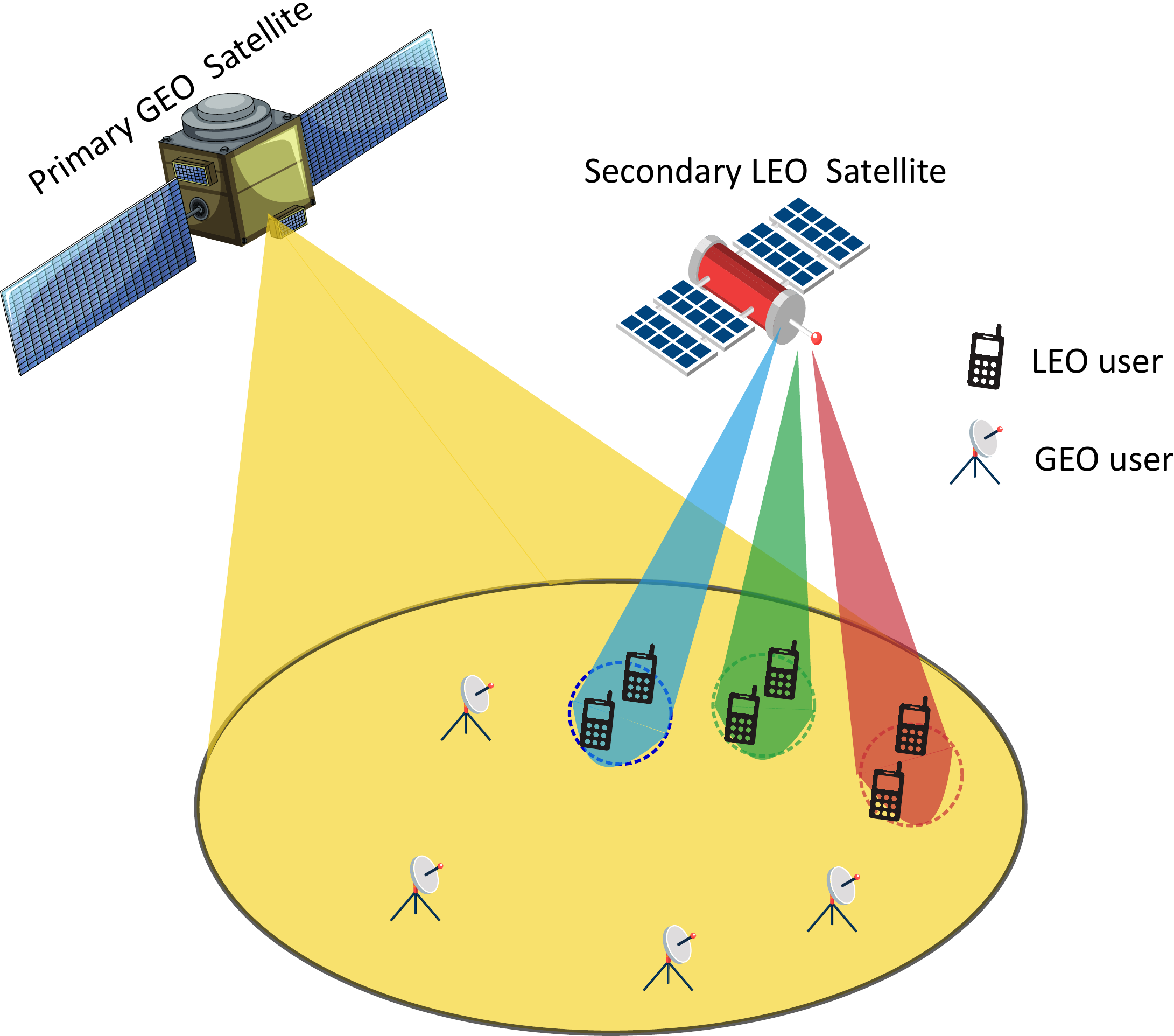}
\caption{System Model of cognitive radio enabled SatCom network using RSMA technique, where a licensed primary GEO SatCom network share the available spectrum resources with unlicensed secondary LEO SatCom network in downlink transmission. }
\label{blocky}
\end{figure*}
%%%%%%%%%%%%%
\section{System Model and Problem Formulation}
As shown in Fig. \ref{blocky}, we consider a cognitive radio-enabled Ka-band multi-beam LEO SatCom network using RSMA technique. In the considered model, LEO SatCom operates as a secondary network in the coverage area of primary GEO SatCom network by sharing the same spectrum resources. LEO SatCom and GEO SatCom share $K$ resource blocks for accommodating single antenna ground users. To guarantee the quality of services of primary GEO SatCom, our proposed optimization framework restricts the interference temperature from LEO SatCom to the ground users of primary GEO SatCom. We consider that primary GEO SatCom communicates with ground users using OMA technique such that each resource block can accommodate one ground user. Meanwhile, the secondary LEO SatCom solely reuses the same resource block such that each beam accommodates multiple ground users through a single resource block using RSMA protocol\footnote{In this work, we adopt the 1-layer rate splitting which is expected to be a more practical RSMA scheme \cite{9831440}. Moreover, it is the basic building block of all existing RSMA schemes.}. This work assumes that the channel state information is perfectly known for the entire network. We denote the set of total LEO beams as $M$, total ground users as $U$ and the subset of users associated with $m$ beam via $k$ resource block as $U_m$, where $U_m<<U$ and $m\in M$. Let's call the total amount of antenna feeds $N_t$. According to the authors in \cite{yin2020rate}, the array-fed reflector can convert $N_t$ feed signals into $M$ transmitted signals (i.e., one signal per beam) that can be broadcast across the multi-beam coverage region. Next, considering the modern satellite architecture \cite{wang2019multicast}, the proposed system considers a single feed per beam, i.e., only one feed is required to generate one beam. Based on this observation, the number of antenna feeds is similar to the number of beams. The overall messages intended to ground users associated with $M$ beams over $K$ resource blocks can be expressed as $ W_{1,1},\dots,W_{m,k},\dots, W_{M,K}$. Each message splits into two parts, i.e., private message and common message such as $W_{m,k}\rightarrow\{W_{m,c,k},W_{m,u,k}\}$. All the common messages are combined and encoded into a single common stream shared by all group such as $W_{1,c,1},\dots,W_{m,c,k},\dots, W_{M,c,K}$. Accordingly, the private messages are combined and encoded into private stream over each beam independently, i.e., $W_{m,1,k},\dots,W_{m,u,k},\dots,W_{m,U_m,k}$.

The single common message and $U_m$ private messages are independently encoded into stream $s_{m,c,k}, s_{m,1,k},\dots, s_{m,u,k}, \dots,s_{m,U_m,k}$, where $s_{m,c,k}$ and $s_{m,u,k}$ are the encoded private and common symbols. Next we define a linear precoding vectors as ${\bf w}_{m,c,k}, {\bf w}_{m,1,k},\dots,{\bf w}_{m,U_m,k}\in \mathbb{C}_{N_t\times1}$ and $\bf{W}=[{\bf w}_{m,c,k}, {\bf w}_{m,1,k},\dots,{\bf w}_{m,U_m,k}]\in\mathbb{C}^{N_t\times(U_m+1)}$. The overall transmitted signal of $m$ beam to $U_m$ users over $k$ resource block can be written as
\begin{align}
s_{m,k}=\sqrt{\eta_{m,c,k}p_{m,k}}{\bf w}_{m,c,k}s_{m,c,k}+\sum\limits_{u=1}^{U_m}\sqrt{\eta_{m,u,k}p_{m,k}}{\bf w}_{m,u,k}s_{m,u,k},
\end{align}
where $p_{m,k}$ denotes the transmit power of $m$ beam at $k$ resource block. $\eta_{m,0,k}$ and $\eta_{m,u,k}$ are the power allocation coefficients of the common message $s_{m,0,k}$ and private message $s_{m,u,k}$ over $k$ resource block. In practice, a satellite requires very high bandwidth for communications, the channel between the satellite and the ground terminal can be considered to be block fading, which remains constant over a block of the enormous number of symbols. Based on this observation and following research work in \cite{9512414}, we consider a block faded channel model from $m$ beam to $u$ user over $k$ resource as ${\bf h}_{m,u,k}=\bar h_{m,u,k}e^{\bar j2\pi\xi_{m,u,k}}\in\mathbb{C}^{N_t\times1}$, where $\xi_{m,u,k}$ denotes the Doppler shift, $\bar j=\sqrt{-1}$ \cite{you2020massive11}, and $\bar h_{m,u,k}$ represents the complex-valued channel gain. Considering large and small scale fading, $\bar h_{m,u,k}$ can be expressed as
\begin{align}
\bar h_{m,u,k}=\sqrt{G_TG_R \Big(\frac{c}{4\pi f_c d_{m,u,k}}\Big)^2}
\end{align}
where $G_R$ is the gains of received antenna while $c$ is the speed of light and $f_c$ is the carrier frequency. Moreover, $d_{m,u,k}$ denotes the distance between $m$ beam and $u$ user over $k$ resource block. Furthermore, $G_{m,u,k}$ is the transmit antenna gain $m$ beam which mostly dependant on antenna radiation pattern and user locations. In the proposed system, it is approximated as
\begin{align}
 G_{m,u,k} =   G_{max}\left[\frac{J_1(\varrho_{m,u,k})}{2\varrho_{m,u,k}}+36\frac{J_3(\varrho_{m,u,k})}{\varrho^3_{m,u,k}}\right]^2,
\end{align}
where $\varrho_{m,u,k}=2.07123\sin{(\theta_{m,u,k})}/\sin(\theta_{3dB})$, where $\theta_{m,u,k}$ is the angle between ground user and the centre of $m$ beam for any given ground user location. Moreover, the angle of 3 dB loss relative to the beam's center is denoted by $\theta_{3dB}$. Furthermore, $G_{max}$ is the maximum gain observed at each beam centre. In addition, $J_1$ and $J_2$ are the first-kind
Bessel functions with order 1 and order 3, respectively. The total signal that $u$ ground user receives from $m$ beam over $k$ resource block can be expressed as
\begin{align}
{\bf y}_{m,u,k}&={\bf h}_{m,u,k}\sqrt{p_{m,k}\eta_{m,0,k}}s_{m,0,k}\nonumber\\&+{\bf h}_{m,u,k}\sum\limits_{i=1}^{U_m}\sqrt{\eta_{m,i,k}p_{m,k}}s_{m,i,k}\nonumber\\& +{\bf g}^{m'}_{m,u,k}\sqrt{q_{m',u',k}},z_{m',u',k}+\omega_{m,u,k},\label{3}
\end{align}  
where ${\bf g}^{m'}_{m,u,k}$ is the channel gain from the $m'$ GEO\footnote{The channel model of GEO is different from LEO due to characteristic differences in orbits and terminals. The main difference is that we consider Doppler shift effect in the channel model of LEO due to very high velocity. This effect can be safely ignored in the channel model of GEO.} beam to $u$ user of $m$ LEO beam over $k$ resource block \cite{khan2022ris}, $q_{m',u',k'}$ represents transmit power of $m'$ beam for $u'$ user over $k$ resource block, and $z_{m',u',k'}$ is the transmitted signal. Moreover, $\omega_{m,u,k}\in\mathcal{CN}(0,N_0\boldsymbol{I}_{N_t})$ is the additive white Gaussian noise of $u$ user. To ensure that the common signal can be successfully
decoded by all users associated with $m$ beam of LEO satellite over $k$ resource block, the achievable data rate of the common message can be stated as
\begin{align}
R_{m,c,k}= \underset{u\in U}{\min} B\log_2(1+\gamma_{m,c,k}),
\end{align} 
with 
\begin{align}
\gamma_{m,c,k}=\frac{|{\bf w}^H_{m,c,k}{\bf h}_{m,u,k}|^2\eta_{m,0,k}p_{m,k}}{|{\bf g}^{m'}_{m,u,k}|^2q_{m',u',k}+I_{ R_{m,c,k}}+\sigma^2},
\end{align}
where $I_{ R_{m,c,k}}=|{\bf w}^H_{m,u,k}{\bf h}_{m,u,k}|^2\sum_{j=1, j\neq u}^{U} x_{m,j,k}\eta_{m,j,k}p_{m,k}$ is the interference of RSMA users during decoding of common signal, $B$ is the bandwidth available at $m$ beam and $\sigma^2={\bf w}^H_{m,c,k}\omega_{m,u,k}$. Next we define an efficient resource block user assignment to beam criterion based on binary decision. Let us consider that $x_{m,u,k}\in\{0,1\}$ is the binary variable of resource block user assignment which can be expressed as
\begin{align}
x_{m,u,k}=\left\{ 
  \begin{array}{ c l }
    1 & \quad \textrm{if $u$ user is assigned to $m$ beam over $k$} \\
    &\quad\textrm{resource block},\\
    0                 & \quad \textrm{otherwise}.
  \end{array}
\right.
\end{align}

Since $R_{m,c,k}$ is the shared signal between users on $k$ resource block such that $ C_{m,u,k}$ denotes the portion of $u$'s user data rate, where $\sum_{u=1}^{U} C_{m,u,k}\leq  R_{m,c,k}$.

After successfully decoding the common signal $s_{m,0,k}$, each user also decodes its private signal, the achievable data rate of $u$ user to decode its private signal $s_{m,u,k}$ can be written as
\begin{align}
R_{m,u,k}= B\log_2(1+\gamma_{m,u,k}),
\end{align}
with
\begin{align}
\gamma_{m,u,k}=\frac{|{\bf w}^H_{m,u,k}h_{m,u,k}|^2\eta_{m,u,k}p_{m,k}}{g^{m'}_{m,u,k}q_{m',u',k}+I_{ R_{m,u,k}}+\sigma^2},
\end{align}
where $I_{ R_{m,u,k}}=|{\bf w}^H_{m,u,k}{\bf h}_{m,u,k}|^2\sum_{j=1,j\neq u}^{U}x_{m,j,k}\eta_{m,j,k}p_{m,k}$ denotes the RSMA interference during decoding of private signal. Given the achievable data rate of common and private signals, the total achievable data rate of $u$ user from $m$ beam over $k$ resource block can be written as $ R_{tot}=  C_{m,u,k}+ R_{m,u,k}$. 

To ensure the quality of services of GEO SatCom, our optimization framework restrict the interference temperature from LEO SatCom to GEO ground users. Specifically, the interference from $m$ beam of LEO satellite over $k$ resource block at GEO user should be restricted as 
\begin{align}
f^m_{m',u',k} p_{m,k}\leq I_{th}, \forall m,k
\end{align}
where $I_{th}$ is the maximum interference temperature threshold to GEO user from $m$ beam of LEO satellite over $k$ resource block.

Given the proposed system model, we seek to maximize the spectral efficiency of cognitive radio enabled LEO SatCom network by optimizing the transmit power of all beams, RSMA power allocation over each beam and resource block user assignment while ensuring the quality of services of both LEO ground users and GEO ground users. This optimal framework can be achieved by formulating and solving the following sum rate maximization problem
\begin{align}
R_{sum}=\sum\limits_{m=1}^M\sum\limits_{k=1}^K\sum\limits_{u=1}^{U}x_{m,u,k} (C_{m,u,k}(\boldsymbol{\eta},{\bf c})+ R_{m,u,k}(\boldsymbol{\eta}))
\end{align}
where $\boldsymbol{\eta}=[\eta_{m,0,k},\eta_{m,1,k},\eta_{m,2,k},\dots,\eta_{m,u,k},\dots,\eta_{m,U,k}]$ is the vector of power allocation coefficients and ${\bf c}=[ C_{m,1,k}, C_{m,2,k},\dots,C_{m,u,k},\dots, C_{m,U,k}]$ denotes the common data rate vector of all users over $k$ subcarrier. The maximum sum rate can be achieved through efficient subcarrier beam assignment and power allocation at secondary LEO satellite while control the interference temperature to primary GEO satellite and guarantee the minimum data rate of LEO users. A complete optimization framework of joint power allocation and resource block user assignment can be modeled as
%%%%%%%%%%%%%%%%%%%%%%%%%%%
\begin{alignat}{2}
& \underset{{(\boldsymbol{\eta},{\bf c},{\bf x},{\bf p}, {\bf w})}}{\text{max}}\  R_{sum}\label{7}\\
s.t.&
\begin{cases}
 \mathcal C_1: \sum\limits_{m=1}^M\sum\limits_{k=1}^Kx_{m,u,k}(C_{m,u,k}+ R_{m,u,k})\geq  R_{min},\ \forall u, \\
 \mathcal C_2: \sum\limits_{u=1}^{U}x_{m,u,k} C_{m,u,k}\leq  R_{m,c,k}, \forall m,k, \\
 \mathcal C_3: f^{m}_{m',u',k}p_{m,k}\leq I_{th},\forall m,k,\\
 \mathcal C_4: \eta_{m,0,k}+\sum\limits_{u=1}^{U}x_{m,u,k}\eta_{m,u,k}\leq 1,\ \forall m,k, \\
 \mathcal C_5: \sum\limits_{m=1}^M\sum\limits_{k=1}^Kp_{m,k}\leq P_{tot}, \\
 \mathcal C_6: |{\bf w}_{m,u,k}|^2=1, \forall m,k,u,\nonumber\\
 \mathcal C_7: \sum\limits_{m=1}^M\sum\limits_{k=1}^K x_{m,u,k}=1, \forall u, \nonumber \\
 \mathcal C_8: x_{m,u,k}\in\{0,1\}, \forall m,k,u, \nonumber \\
\end{cases}
\end{alignat}
%%%%%%%%%%%%%%%%%%%%%%%%%%
 where constraint $\mathcal C_1$ guarantees the data rate of $u$ user over $k$ resource block and $R_{min}$ denotes the threshold of minimum data rate. Constraint $\mathcal C_2$ bounds the proposed framework to successfully decode the common signal of all users associated with $m$ beam over $k$ resource block. Constraint $\mathcal C_3$ limits the interference temperature from $m$ beam of LEO satellite to $u'$ GEO user over $k$ resource block. Constraint $\mathcal C_4$ control the total allocated power at each beam while constraint $\mathcal C_5$ controls the total energy consumption of LEO satellite, where $P_{tot}$ shows the total power budget threshold. Moreover, $\mathcal C_6$ represents the precoding vector constraint. Then, Constraints $\mathcal C_7$ and $\mathcal C_8$ say that a user should be assigned a single beam and only one resource block.
  
\section{Proposed Optimization Solution}
It can be observed that the sum rate maximization problem in (\ref{7}) is non-convex due the rate expressions and binary variable. Moreover, the problem is coupled on multiple optimization variables and poses high complexity. Based on the nature of this problem, it is very challenging to obtain the joint optimal solution. To reduce the problem complexity and make the problem more tractable, we obtain a suboptimal yet efficient solution. For any given resource allocation at the system, the signal to interference plus noise ratios of both private and common messages depend on the precoding vectors. Following the work in \cite{9369326}, the efficient precoding vectors which balance both interference and noise can be expressed as
\begin{align}
{\bf w_{m,c,k}}=(N_0\boldsymbol{I}_{N_t}+{\bf h}_{m,u,k}{\bf h}^H_{m,u,k}\sum_{j=1}^{U}x_{m,j,k}\eta_{m,j,k}p_{m,k})^{-1}{\bf h}_{m,u,k}, \forall c
\end{align}
\begin{align}
{\bf w_{m,u,k}}=(N_0\boldsymbol{I}_{N_t}+{\bf h}_{m,u,k}{\bf h}^H_{m,u,k}\sum_{j\neq u}^{U}x_{m,j,k}\eta_{m,j,k}p_{m,k})^{-1}{\bf h}_{m,u,k}, \forall p
\end{align}     
Next the efficient solution of power allocation and resource block user assignment can be achieved in three steps. First, we adopt successive convex approximation (SCA) technique to reduce the complexity and make the optimization problem more tractable. Second, for any given resource block user assignment, the efficient transmit power can be calculated using KKT conditions. Third, using the calculated value of transmit power, we design efficient resource block user assignment based on greedy approach. According to SCA technique \cite{7296696}, the original non-convex functions can be efficiently replaced by properly chosen surrogates. By applying this, the data rate of $u$ user associated with $m$ beam over $k$ resource block can be written as
\begin{align}
R_{m,u,k}= W \tau_{m,u,k} \log_2(\gamma_{m,u,k})+\varpi_{m,u,k},
\end{align}
where $\tau_{m,u,k}=\frac{\gamma_{m,u,k}}{1+\gamma_{m,u,k}}$ and $\varpi_{m,u,k}=\log_2(1+\gamma_{m,u,k})-\tau_{m,u,k} \log_2(\gamma_{m,u,k})$. Similarly, we apply SCA for the data rate of the common message as
\begin{align}
R_{m,c,k}= \underset{u\in U}{\min} W\log_2(1+\gamma_{m,c,k})+\varpi_{m,c,k},
\end{align}
where $\tau_{m,c,k}=\frac{\gamma_{m,c,k}}{1+\gamma_{m,c,k}}$ and $\varpi_{m,c,k}=\log_2(1+\gamma_{m,c,k})-\tau_{m,c,k} \log_2(\gamma_{m,c,k})$. Next we calculate efficient transmit power for any given resource block user assignment.

\subsection{Efficient Power Allocation}
For any given resource block user assignment, we calculate efficient power budget for different beams at LEO satellite and RSMA power allocation for users at each beam. The original joint problem in (\ref{7}) can be efficiently re-transformed to power allocation subproblem as
%%%%%%%%%%%%%%%%%%%%%%%%%%%
\begin{alignat}{2}
P_1 \quad & \underset{{(\boldsymbol{\eta},{\bf c},{\bf p})}}{\text{max}}\  R_{sum}\label{p1}\\
s.t.&\quad \mathcal C_1,\mathcal C_2,\mathcal C_3,\mathcal C_4,\mathcal C_5. \nonumber
\end{alignat}
%%%%%%%%%%%%%%%%%%%%%%%%%%
Now we define a Lagrangian of problem (\ref{p1}) as
\begin{align}
    & L=-\sum\limits_{m=1}^M\sum\limits_{k=1}^K\sum\limits_{u=1}^{U}x_{m,u,k} (C_{m,u,k}(\boldsymbol{\eta},{\bf c})+ R_{m,u,k}(\boldsymbol{\eta}))+\nonumber\\&\sum\limits_{u=1}^{U}\lambda1_{u}\Big( R_{min}-\sum\limits_{m=1}^M\sum\limits_{k=1}^Kx_{m,u,k}(C_{m,u,k}+ R_{m,u,k})\Big)+\nonumber\\&\sum\limits_{m=1}^M\sum\limits_{k=1}^K\lambda2_{m,k}x_{m,u,k}\Big(\sum\limits_{u=1}^{U} C_{m,u,k}-  R_{m,c,k}\Big)+\sum\limits_{m=1}^M\sum\limits_{k=1}^K\nonumber\\&\lambda3_{m,k}\Big(f^m_{m',u',k} p_{m,k}- I_{th}\Big)+\sum\limits_{m=1}^M\sum\limits_{k=1}^K\lambda4_{m,k}\Big(\eta_{m,0,k}+\nonumber\\&\sum\limits_{u=1}^{U}x_{m,u,k}\eta_{m,u,k}- 1\Big)+\lambda5\Big(\sum\limits_{m=1}^M\sum\limits_{k=1}^Kp_{m,k}- P_{tot}\Big).
\end{align}
where $\boldsymbol{\lambda}$ represent Lagrangian multipliers. Now we applying KKT conditions \cite{khan2022energy} to calculate transmit power over each beam by computing partial derivation with respect to $p_{m,k}$. Applying KKT conditions we know that:
\begin{align}
    \dfrac{\partial L}{\partial p_{m,k}}=0,
\end{align}
Computing the gradient we get:
\begin{align}
    &\sum\limits_{u=1}^U \sum_{j=1,j\neq u}^{U} \Bigg(\lambda5+f_{m',u',k}^m \lambda3_{m,k} x_{m,u,k}-\dfrac{\lambda2_{m,k} \gamma_{m,c,k} W x_{m,u,k} }{p_{m,k}}-\dfrac{\lambda1_u \gamma_{m,u,k W x_{m,u,k}}}{p_{m,k}}-\nonumber\\&\dfrac{(I_p+\sigma^2)\gamma_{m,u,k} W x_{m,u,k}}{p_{m,k} (I_p \sigma^2 +h_{m,u,k}\eta_{m,j,k} p_{m,k}x_{m,j,k})}-\dfrac{(I_p+\sigma^2)\gamma_{m,j,k} W x_{m,j,k}}{p_{m,k} (I_p \sigma^2 +h_{m,j,k}\eta_{m,u,k} p_{m,k}x_{m,u,k})}   \Bigg)=0,
\end{align}
Simplifying this we get:
\begin{align}
    &\sum\limits_{u=1}^U \sum_{j=1,j\neq u}^{U} \Bigg(-(I_p+\sigma^2)\gamma_{m,j,k} W x_{m,j,k}(I_p+\sigma^2+h_{m,u,k}\eta_{m,j,k}p_{m,k}x_{m,j,k})-(I_p+\sigma^2)\gamma_{m,u,k} \nonumber\\&W x_{m,u,k}(I_p+\sigma^2+h_{m,j,k}\eta_{m,u,k}p_{m,k}x_{m,u,k})+p_{m,k}(I_p+\sigma^2+h_{m,u,k}\eta_{m,j,k}p_{m,k}x_{m,j,k})(\lambda5+\nonumber\\& f_{m',u',k}^m \lambda3_{m,k}x_{m,u,k})(I_p+\sigma^2+h_{m,j,k}\eta_{m,u,k}p_{m,k}x_{m,u,k})+(I_p+\sigma^2+h_{m,u,k}\eta_{m,j,k}p_{m,k}x_{m,j,k})\nonumber\\&(I_p+\sigma^2+h_{m,j,k}\eta_{m,u,k}p_{m,k}x_{m,u,k})(-\lambda2_{m,k}\gamma_{m,c,k} W x_{m,u,k}-\lambda1_n\gamma_{m,u,k}W x_{m,u,k})\Bigg)=0,
\end{align}
Further simplifying and rearranging the equation we can write the equation in terms of $p_{m,k}$ as: 

\begin{align}
    \zeta_3 p_{m,k}^3+\zeta_2 p_{m,k}^2+\zeta_1 p_{m,k}+\zeta_0=0,\label{eq11}
\end{align}
where the values of $\zeta_3$, $\zeta_2$, $\zeta_1$ and $\zeta_0$ are defined as

\begin{align}
    \zeta_3=&\sum\limits_{u=1}^U \sum_{j=1,j\neq u}^{U} h_{m,j,k} h_{m,u,k} \eta_{m,j,k}\eta_{m,u,k} x_{m,j,k} x_{m,u,k}\nonumber\\&(\lambda5+f^{m}_{m',u',k} \lambda3_{m,k}x_{m,u,k}),
\end{align}
\begin{align}
    \zeta_2=& \sum\limits_{u=1}^U \sum_{j=1,j\neq u}^{U} h_{m,j,k} \eta_{m,u,k} (I_p+\sigma^2) x_{m,u,k}(\lambda5+f^{m}_{m',u',k} \nonumber\\&\lambda3_{m,k} x_{m,u,k})+h_{m,u,k} \eta_{m,j,k} x_{m,j,k} (\lambda5 (I_p +\sigma^2)+\nonumber\\& f^{m}_{m',u',k}\lambda3_{m,k}(I_p+\sigma^2)x_{m,u,k}-h_{m,j,k} \eta_{m,u,k}\nonumber\\&(\lambda2_{m,k}\gamma_{m,c,k}+\lambda1_n \gamma_{m,u,k}) W x_{m,u,k}),
\end{align}
\begin{align}
    \zeta_1=& \sum\limits_{u=1}^U \sum_{j=1,j\neq u}^{U} (I_p+\sigma^2)(\lambda5 \sigma^2+f_{m,k} \lambda3_{m,k} \sigma^2 x_{m,u,k}\nonumber\\&+I_p(\lambda5+f^{m}_{m',u',k}\lambda3_{m,k}x_{m,u,k})+W (-h_{m,j,k} \eta_{m,u,k}\nonumber\\&(\lambda2_{m,k} \gamma_{m,c,k}+\gamma_{m,u,k}+\lambda1_u\gamma_{m,u,k}) x_{m,u,k}\nonumber\\& -h_{m,u,k}\eta_{m,j,k}x_{m,j,k}(\gamma_{m,j,k} x_{m,j,k}\nonumber\\&+\lambda2_{m,k} \gamma_{m,c,k} x_{m,u,k}+\lambda1_u \gamma_{m,u,k} x_{m,u,k}))),
\end{align}
\begin{align}
    \zeta_0=& \sum\limits_{u=1}^U \sum_{j=1,j\neq u}^{U} -(I_p+\sigma^2)^2 W(\gamma_{m,j,k} x_{m,j,k}+(\lambda2\gamma_{m,c,k}\nonumber\\&+\gamma_{m,u,k}+\lambda1_u\gamma_{m,u,k})x_{m,u,k}),
\end{align}
with $I_p=g^{m'}_{m,u,k}q_{m',u',k}$ which represents the interference from GEO transmissions to LEO ground users. Next we find the solution of $p_{m,k}$ which can be obtained by solving the polynomial in (\ref{eq11}) using any mathematical solver. Then solving for $n_{m,n,k}$  can be obtained as
\begin{align}
    \eta_{m,n,k}= \frac{\mu1 \pm \sqrt{\mu2}}{\mu3},
\end{align}
where there values of $\mu1$, $\mu2$ and $\mu3$ can be written as
\begin{align}
    \mu1=&\sum_{j=1,j\neq u}^{U}-\lambda4_{m,k}(I_p+\sigma^2)+h_{m,j,k} p_{m,k} W(-\gamma_{m,j,k}\nonumber\\& x_{m,j,k}+(1+\lambda1_u)\gamma_{m,u,k}x_{m,u,k}),
\end{align}
\begin{align}
    \mu2=&\sum_{j=1,j\neq u}^{U} 4 h_{m,j,k}(1+\lambda1_u)\lambda4_{m,k} p_{m,k} (I_p+\sigma^2) \gamma_{m,u,k} W\nonumber\\& x_{m,u,k}+(\lambda4_{m,k}(I_p+\sigma^2) h_{m,j,k} p_{m,k}(\gamma_{m,j,k}x_{m,j,k}-\nonumber\\&(1+\lambda1_u)\gamma_{m,u,k}x_{m,u,k}))^2,
\end{align}
\begin{align}
    \mu3=&\sum_{j=1,j\neq u}^{U} 2 h_{m,j,k}\lambda4_{m,k} p_{m,k}x_{m,u,k}.
\end{align}
Similarly we solve $\eta_{m,0,k}$. The value of $\eta_{m,0,k}$ can be computed as
\begin{align}
    \eta_{m,0,k}=\frac{\lambda2_{m,k} \gamma_{m,c,k} W}{\lambda4_{m,k}}.
\end{align}
It can be seen that the considered problem is affine with respect to $ C_{m,u,k}$. We know that for affine problems the local maxima is also the global maxima, as in any given interval there can be only one extrema (problems having multiple extrema are non-convex in nature and an affine problem can never be non-convex). Therefore, we employ gradient ascend optimization technique to find the optimal solution. In this technique, the framework updates the parameter in the increasing direction of the gradient, hence, resulting in a solution that provides the maximum value of the objective function. Based on gradient ascent method, in each iteration, the value of $ C_{m,u,k}$ can be updated as:
\begin{align}
 C_{m,u,k}= C_{m,u,k}+\delta (1+\lambda1_u-\lambda2_{m,k}).
\end{align}
Accordingly, the values of Lagrangian multipliers can be updated as:
\begin{align}
&\lambda1_{u}=\lambda1_{u}+\delta \Big( R_{min}-\sum\limits_{m=1}^M\sum\limits_{k=1}^K x_{m,u,k} (C_{m,u,k}+R_{m,u,k})\Big),
\end{align}
\begin{align}
&\lambda2_{m,k}=\lambda2_{m,k}+\delta x_{m,u,k} \Big(\sum\limits_{u=1}^{U}C_{m,u,k}- R_{m,c,k}\Big),
\end{align}
\begin{align}
&\lambda3_{m,k}=\lambda3_{m,k}+\delta\Big( f^{m}_{m',u',k}p_{m,k}- I_{th}\Big),
\end{align}
\begin{align}
&\lambda4_{m,k}=\lambda4_{m,k}+\delta\Big(\eta_{m,0,k}+\sum\limits_{u=1}^{U}x_{m,u,k}\eta_{m,u,k}- 1\Big),
\end{align}
\begin{align}
&\lambda5=\lambda5+\delta\Big(\sum\limits_{m=1}^M\sum\limits_{k=1}^K p_{m,k}- P_{tot}\Big).
\end{align}
where $\delta$ is the positive step size.
\begin{algorithm}[!t]
\caption{Resource Block User Assignment}
{\bf Step 1:} Initialize all the system parameters\\
{\bf Step 2:} Assignment process
\begin{enumerate}
\item \textbf{Set} $U_m=U_m=\lceil \frac{U}{M} \rceil$, $x$ = zeros(M,U,K)
\item \textbf{for $a=1:U_m$}
\item \quad\textbf{for $b=1:M$}
\item \quad\quad\textbf{for $c=1:K$}
\item \quad\quad\textbf{Find} $e$ such that $h_{m,e,k}=\max( h_{m,:,k})$ (where \newline\indent\quad\quad$ h_{m,:,k}$ are the channel gains of all users not \newline\indent\quad\quad assigned a beam yet)
\item \quad\quad\textbf{Set} $x_{m,u,k}=1$
\item \quad\quad\textbf{Remove} user $e$ from the list of users awaiting\newline \indent\quad\quad subcarrier and beam assignment
\item \quad\quad\textbf{end for}
\item \quad\textbf{end for}
\item \textbf{end for}
\item \textbf{Return} $x_{m,u,k}$
\end{enumerate}
\label{algo1}
\end{algorithm}

%, where users are equally distributed among all the beams in the LEO system
\subsection{Efficient Resource Block User Assignment}
Next, we solve the problem of efficient resource block user assignment. For available power allocation at satellite, the problem (\ref{7}) can be simplified to a resource block user assignment subproblem as
%%%%%%%%%%%%%%%%%%%%%%%%%%%
\begin{alignat}{2}
P_2 \quad & \underset{{(\boldsymbol{x})}}{\text{max}}\  R_{sum}\label{p2}\\
s.t.&\quad \mathcal C_1,\mathcal C_2,\mathcal C_6,\mathcal C_7. \nonumber
\end{alignat}
%%%%%%%%%%%%%%%%%%%%%%%%%%
To solve problem (\ref{p2}), we propose an efficient algorithm based on a greedy approach. According to this algorithm, the problem can be solved such that the best available option for resource block user assignment can be selected at any given moment \cite{8937029}. This algorithm always goes for the best local option to produce the global best result.
For $U$ LEO users in the system, each beam is transmitting data to $U_m$ subset of users where $U_m$ is computed as $U_m=U_m=\lceil \frac{U}{M} \rceil$ the $\lceil \psi \rceil$ function rounds up $\psi$ to the closest integer. Then each beam of every subcarrier is allocated to the user, which has maximum channel gain on the beam. After this, the assigned user is removed from the list of the users that are not allocated a beam yet. Similarly, a user is assigned to every beam. This process is repeated $U_m$ times. At the end, each beam has approximately $U_m$ users, and we get the efficient solution of $x_{m,u,k}$\footnote{Here, it is important to mention that we do not claim this technique to provide the optimal value of $x_{m,u,k}$. However, it can be seen in the results section that our algorithm provides very good performance compared to the case where channels and beams are assigned randomly.}. The detailed steps of the proposed technique are also summarized in Algorithm 1.

\subsection{Complexity Analysis} 
In this subsection, we discuss the computational complexity of the proposed framework (\textit{proposed framework}) and benchmark frameworks (\textit{proposed framework 1} and \textit{proposed framework 2}). More specifically, the \textit{proposed framework} refers to the optimization framework provided in Section III-A and III-B, where we optimize the values of $x_{m,u,k}$, $p_{m,k}$, $\eta_{m,n,k}$ and $ C_{m,u,k}$, respectively. In \textit{benchmark framework 1}, the values of all other variables are optimized where as the available power is distributed equally among all the beams such that the interference threshold in not violated i.e., $p_{m,k}=min\Big(\dfrac{I_{th}}{f_{m,k}},\dfrac{P_{tot}}{MK}\Big)$. In the case of \textit{benchmark framework 2}, all the other variables are optimized but the resource block and users are assigned to each beam randomly. 

In this work, we refer to "complexity" as the number of iterations required for the convergence of the lagrangian multipliers involved in the \textit{proposed framework}. As recalled, the original problem of sum rate maximization has been decoupled into two subproblems and then solved. In particular, the power budget of each beam and the RSMA power allocation of ground users have been calculated for any given resource block user assignment in the solution of subproblem $P_1$. Then, for the given calculated values of transmit power, the efficient resource block user assignment algorithm has been designed in the solution of subproblem $P_2$. The detailed steps of resource block user assignment are also provided in {\bf Algorithm 1}. Therefore, in a given iteration, the complexity of the \textit{proposed framework} for solving subproblem $P_1$ can be expressed as $\mathcal O(2UM)$, where the $U$ is the number ground users and $M$ denotes the number of LEO beams. Similarly, the complexity of the \textit{proposed framework} for solving subproblem $P_2$ can be stated as $\mathcal O(UK^2)$, where $K$ is the set of resource blocks. Now we consider that the total iterations required for the optimization process is $\varPsi$, then the total computational complexity of the \textit{proposed framework} for solving subproblem $P_1$ and subproblem $P_2$ can be written as $\mathcal O\{\varPsi(2UM+UK^2)\}$. Now we discuss the complexity of \textit{proposed framework 1} and \textit{proposed framework 2}, respectively. Based on the optimization variables, the total complexity of the \textit{proposed framework 1} can be given as $\mathcal O\{\varPsi(2U+UK^2)\}$. Accordingly, the total complexity of \textit{proposed framework 2} $\mathcal O\{\varPsi(2UM)\}$.

%%%%%%%%%%%%%%%%
\begin{figure}[!t]
\centering
\includegraphics [width=0.6\textwidth]{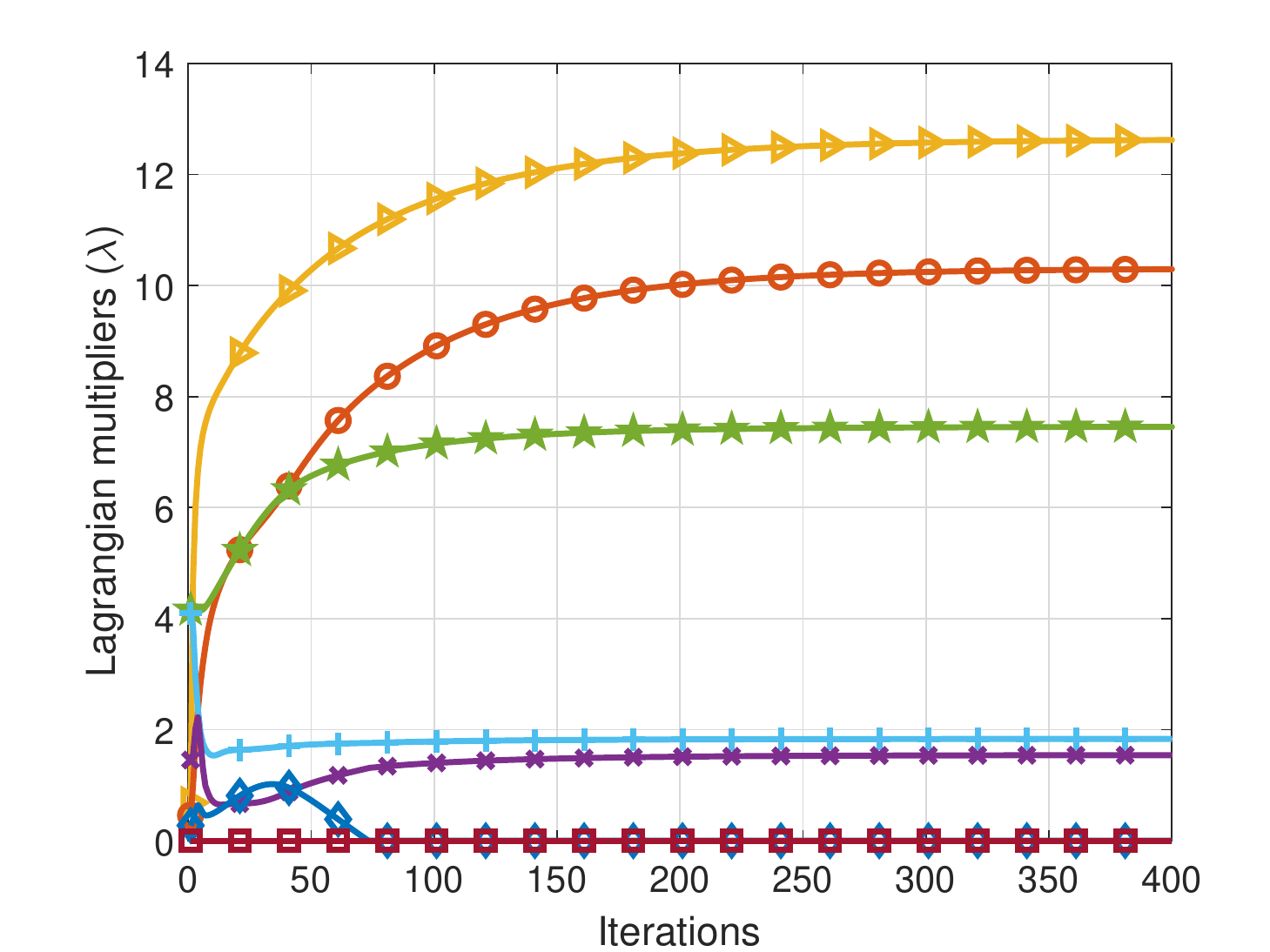}
\caption{Convergence of Lagrangian multipliers involved in the \textit{proposed framework}.}
\label{rf5}
\end{figure}
%%%%%%%%%%%%%
\section{Numerical Results and Discussion}
This section provides numerical results and their discussion. Unless stated otherwise, we follow \cite{9460776,9512414,8352660} to set the values of the following parameters for simulations. It is worth mentioning here that we do not compare the proposed RSMA scheme with NOMA scheme because it has already been proven that a system with NOMA scheme has been significantly outperformed by the system with RSMA scheme \cite{mao2018rate}. The number of resource blocks as $K=5$, number of beams as $M=5$, number of LEO users as $U=2K$, maximum interference threshold to GEO users as $I_{th}=3$ Watts, minimum data rate of each user as $R_{min}=1$ Mbps, total transmit power as $P_{tot}=60$ Watts, $\frac{G_TG_R\lambda^2}{4\pi d_{m,u,k}}=1$ \cite{9419053}, and interference from GEO satellite over each resource block as $I_p=4$ Watts\footnote{In the proposed framework, each beam can use a single resource block for communication with two LEO RSMA users}. Moreover, we consider the frequency band as 19 GHz(Ka) and bandwidth over each beam as $W=10$ MHz. We compare the performance of three optimization frameworks, i.e., \textit{proposed framework}, \textit{benchmark framework 1} and \textit{benchmark framework 2}, respectively. The \textit{proposed framework} is provided in Section III-A and B while \textit{proposed framework 1} and \textit{proposed framework 2} are explained in Section III-C.
%%%%%%%%%%%%%%%%%%%%%%%%%%%%%%%%%%%%%%%%%%%%%%%%%
\begin{table}[!t]
\centering
\caption{Simulation parameters}
\begin{tabular}{|c||c|} 
\hline 
Parameter & Value  \\
\hline\hline
Frequency band $f_{c}$ & 19 GHz(Ka) \\\hline
Total power budget $P_{tot}$ & 60 Watts \\\hline
Number of LEO beams $M$ & 5 \\\hline
Number of LEO ground users $U$ & $2M$ \\\hline
Number of resource blocks $K$ & 5 \\\hline
Interference from LEO to GEO $I_{th}$ & 3 Watts  \\\hline
Interference from GEO to LEO $I{P}$ & 4 Watts \\\hline
Minimum data rate of LEO user $R_{min}$ & 1 Mbps\\\hline
Bandwidth over each LEO beam & 10 MHz\\\hline
Channel realization & $10^3$ \\\hline
Noise power density $\sigma^2$ & -170 dBm \\
\hline 
\end{tabular}
\end{table}
%%%%%%%%%%%%%%%%%%%%%%%%%%%%%%%%%%%%%%%%%%%%%%%%
%%%%%%%%%%%%%%%%
\begin{figure}[!t]
\centering
\includegraphics [width=0.6\textwidth]{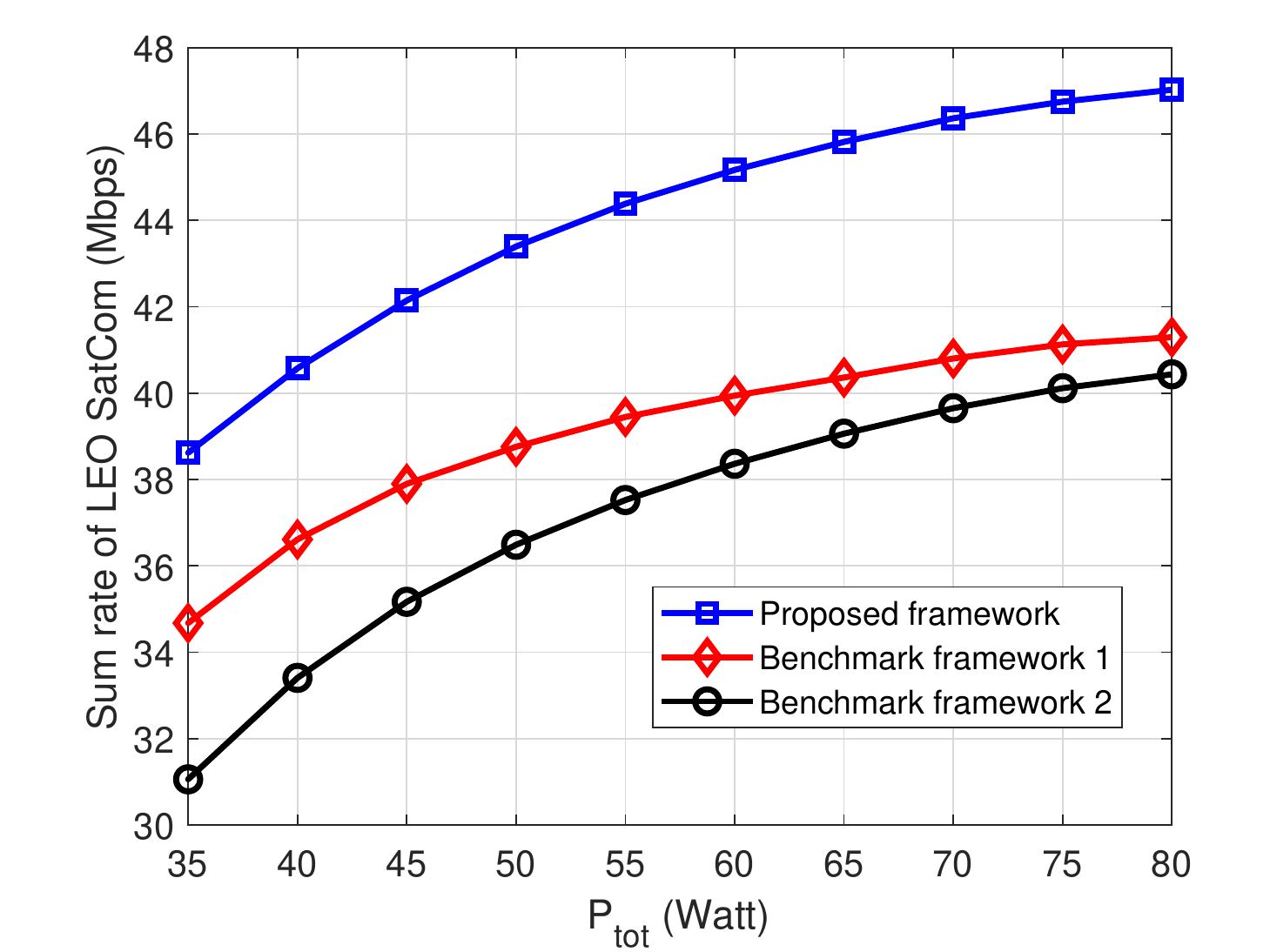}
\caption{Impact of $P_{tot}$ on the sum rate of the system and the performance comparison of different optimization frameworks.}
\label{rf2}
\end{figure}
%%%%%%%%%%%%%

To validate the complexity analysis as discussed in Section III-C, we first plot the vector pf Lagrangian multipliers, i.e., $\boldsymbol{\lambda}$ against the number of iterations as shown in Fig. \ref{rf5}. We can see that different values in $\boldsymbol{\lambda}$ converge within reasonable number of iterations. Overall our \textit{proposed framework} provides significant performance in terms of the sum rate of the system by optimizing multiple variables of the system with reasonable complexity.  

%%%%%%%%%%%%%%%%
\begin{figure}[!t]
\centering
\includegraphics [width=0.6\textwidth]{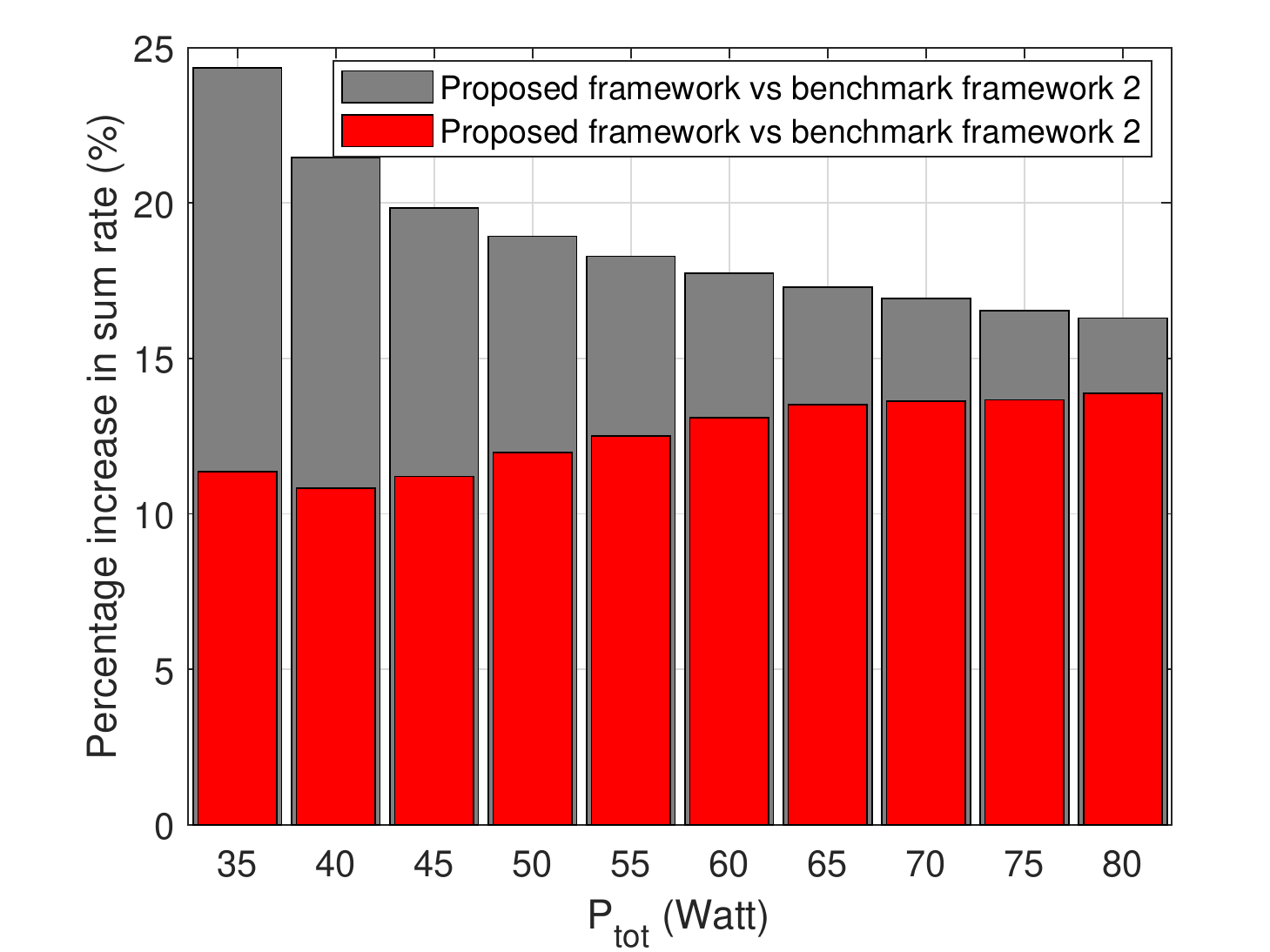}
\caption{Percentage gain in sum rate by the proposed framework compared to benchmark framework 1 and benchmark 2.}
\label{rf11}
\end{figure}
%%%%%%%%%%%%%

Figure \ref{rf2} studies the effect of $P_{tot}$ on the sum rate offered by all the proposed frameworks. An increase in $P_{tot}$ results in increasing the sum rate in all proposed optimization frameworks. An interesting thing to note here is that when $P_{tot}$ is increased, the performance gap between \textit{benchmark framework 2} and \textit{benchmark framework 1} decreases while the gap between \textit{proposed framework} and \textit{benchmark framework 1} increases. This is because in the case of \textit{proposed framework} and \textit{benchmark framework 2}, when the interference threshold of a beam is met with equality, the remaining power is distributed efficiently among other beams, which can not be done in the case of \textit{benchmark framework 1}. It is also clear from Fig. \ref{rf2} that the \textit{proposed framework} provides the best performance for any value of $P_{tot}$. Further, it can be seen that optimizing $x_{m,u,k}$ is more beneficial than optimizing $p_{m,k}$, as \textit{benchmark framework 1} outperforms \textit{benchmark framework 2}. Further, Fig \ref{rf11} shows the percentage gain in sum rate of LEO SatCom network when the \textit{proposed framework} is employed to the proposed model instead for the \textit{benchmark framework 1} and \textit{benchmark framework 2}. This figure shows that the percentage gain in sum rate\footnote{The percentage gain is computed as $\dfrac{(\text{Rate of Proposed framework} - \text{Rate of Benchmark})\times 100}{\text{Rate of Benchmark}}$.} decreases with increasing $P_{tot}$ when the \textit{proposed framework} is compared with \textit{benchmark framework 2}. However, when the \textit{proposed framework} is compared to the \textit{benchmark framework 1}, initially there is a slight decrease in the percentage gain, but after a certain point, as shown in Fig. \ref{rf2}, the \textit{benchmark framework 1} starts to show convergence behaviour where increasing $P_{tot}$ has very little impact on the sum rate of the system. Therefore, at these points, the percentage increase in the sum rate of employing the \textit{proposed framework} starts to increase.  

%%%%%%%%%%%%%%%%
\begin{figure}[!t]
\centering
\includegraphics [width=0.6\textwidth]{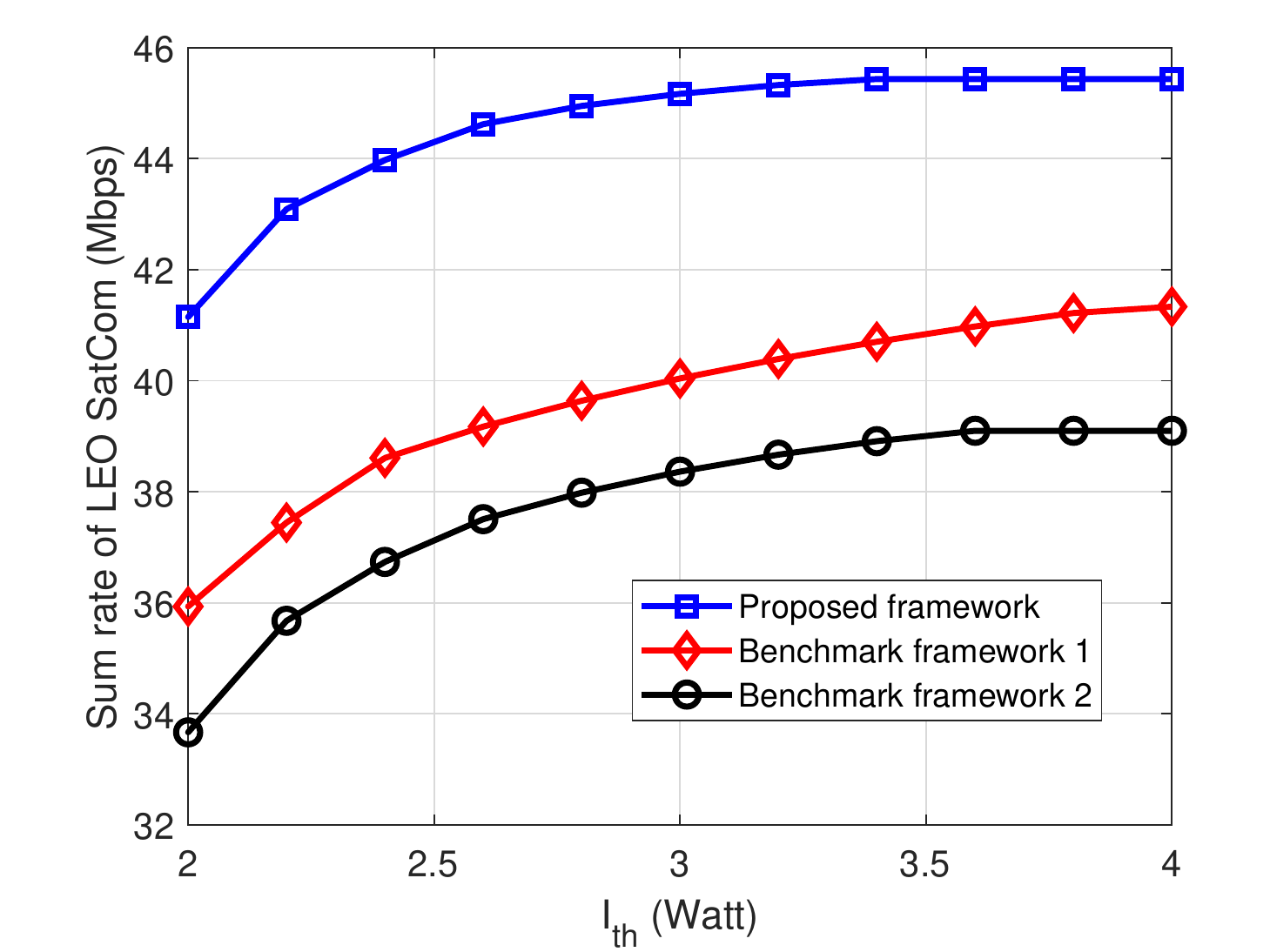}
\caption{The effect of increasing $I_{th}$ on the sum rate of the system offered by all schemes for increase the total available power of LEO SatCom network.}
\label{rf3}
\end{figure}
%%%%%%%%%%%%%
%%%%%%%%%%%%%%%%
\begin{figure}[!t]
\centering
\includegraphics [width=0.6\textwidth]{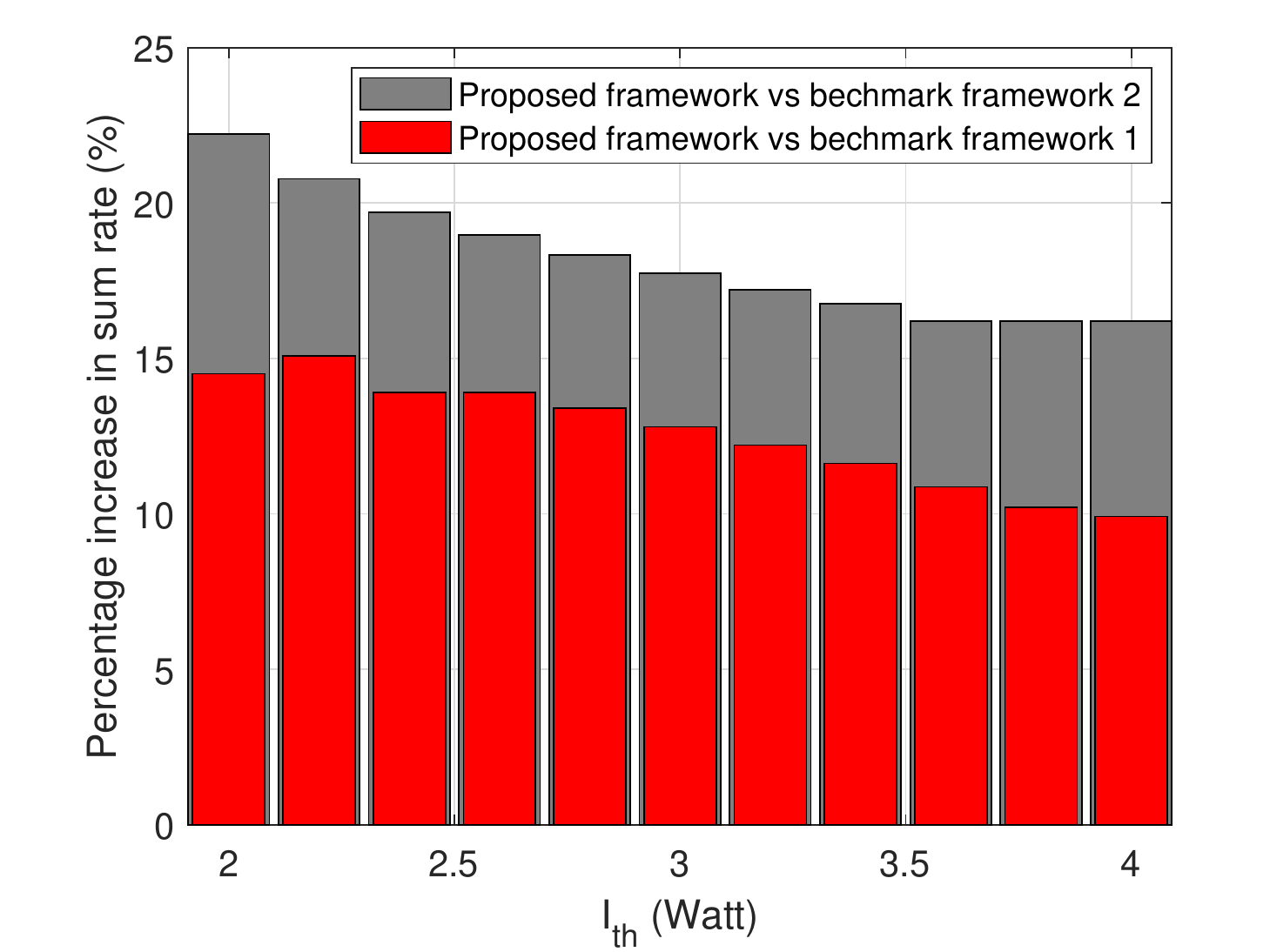}
\caption{Percentage gain in sum rate by the proposed framework compared to benchmark framework 1 and benchmark 2 when increase the interference temperature threshold from LEO SatCom to GEO SatCom.}
\label{rf22}
\end{figure}
%%%%%%%%%%%%%
%tage gain in %%%%%%%%%%%cen%%%%
\begin{figure}[!t]
\centering
\includegraphics [width=0.6\textwidth]{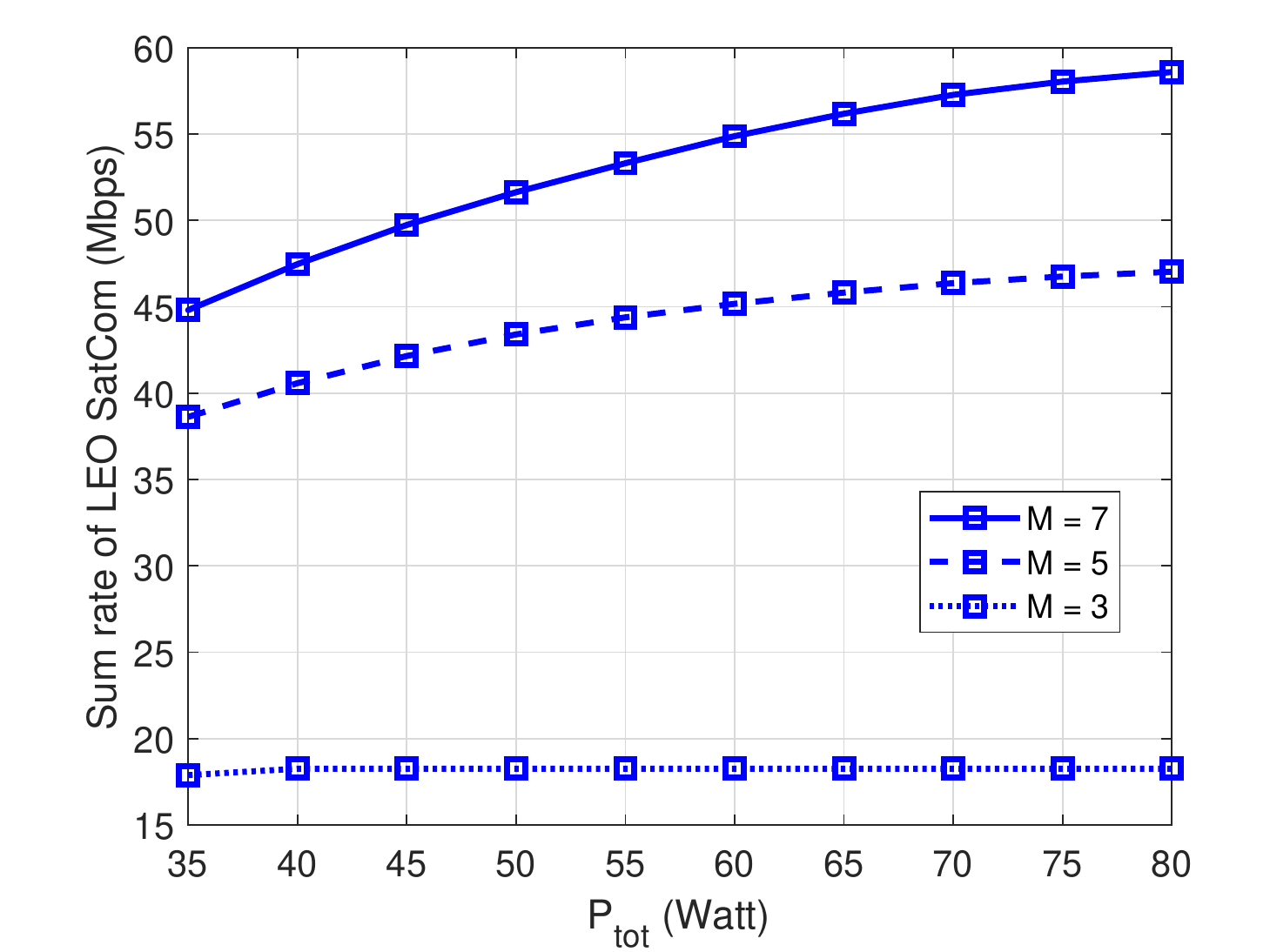}
\caption{Impact of number of beams at LEO satellite on the performance of the system when increasing the total transmit power.}
\label{rf4}
\end{figure}
%%%%%%%%%%%%%

The impact of increasing $I_{th}$ on the sum rate of the system is shown in Fig. \ref{rf3}. An increase in $I_{th}$ results in increasing the sum rate of LEO SatCom network because the transmission power can be increased while satisfying the interference threshold. However, after a certain value of $I_{th}$ when the threshold is further increased, the sum rate remains unchanged in the case of \textit{proposed framework}. Because at this point, the transmission is already being done with full available power. An interesting thing to note in Fig. \ref{rf3} is that initially, the gap between \textit{benchmark framework 1} and \textit{benchmark framework 2} is more, which decreases with increasing $I_{th}$ but after a certain point, the gap starts to increase again. This is because at smaller values of $I_{th}$, the transmission power of all the beams is bounded by the interference threshold. When the value of $I_{th}$ is increased, the transmission power of some beams becomes unbounded by the threshold. At these points, the frameworks where $p_{m,k}$ is optimized allocate the extra power from the bounded beams to other beams. Thus, the gap in performance increases. However, after a certain point, when $I_{th}$ is further increased, the transmission power of no beam is bounded by $I_{th}$. Hence, at these points, the benefit of optimizing $p_{m,k}$ increases. Therefore, the gap between \textit{benchmark framework 1} and \textit{benchmark framework 2} increases and performance gap between \textit{benchmark framework 1} and \textit{proposed framework} decreases at these values of $I_{th}$. Moreover, Fig \ref{rf22} depicts the percentage gain in sum rate of LEO SatCom network when the \textit{proposed framework} is employed to the proposed model instead for the \textit{benchmark framework 1} and \textit{benchmark framework 2}. This figure shows that the percentage gain in sum rate decreases with increasing $I_{th}$ when the \textit{proposed framework} is compared with \textit{benchmark framework 2}. However, when the \textit{proposed framework} is compared to the \textit{benchmark framework 1}, initially there is a slight decrease in the percentage gain, but after a certain point, as shown in Fig. \ref{rf2}, the \textit{benchmark framework 1} starts to show convergence behaviour where increasing $I_{th}$ has very little impact on the sum rate of the system. Therefore, at these points, the percentage increase in the sum rate of employing the \textit{proposed framework} starts to increase.

\begin{figure}[!t]
\centering
\includegraphics [width=0.6\textwidth]{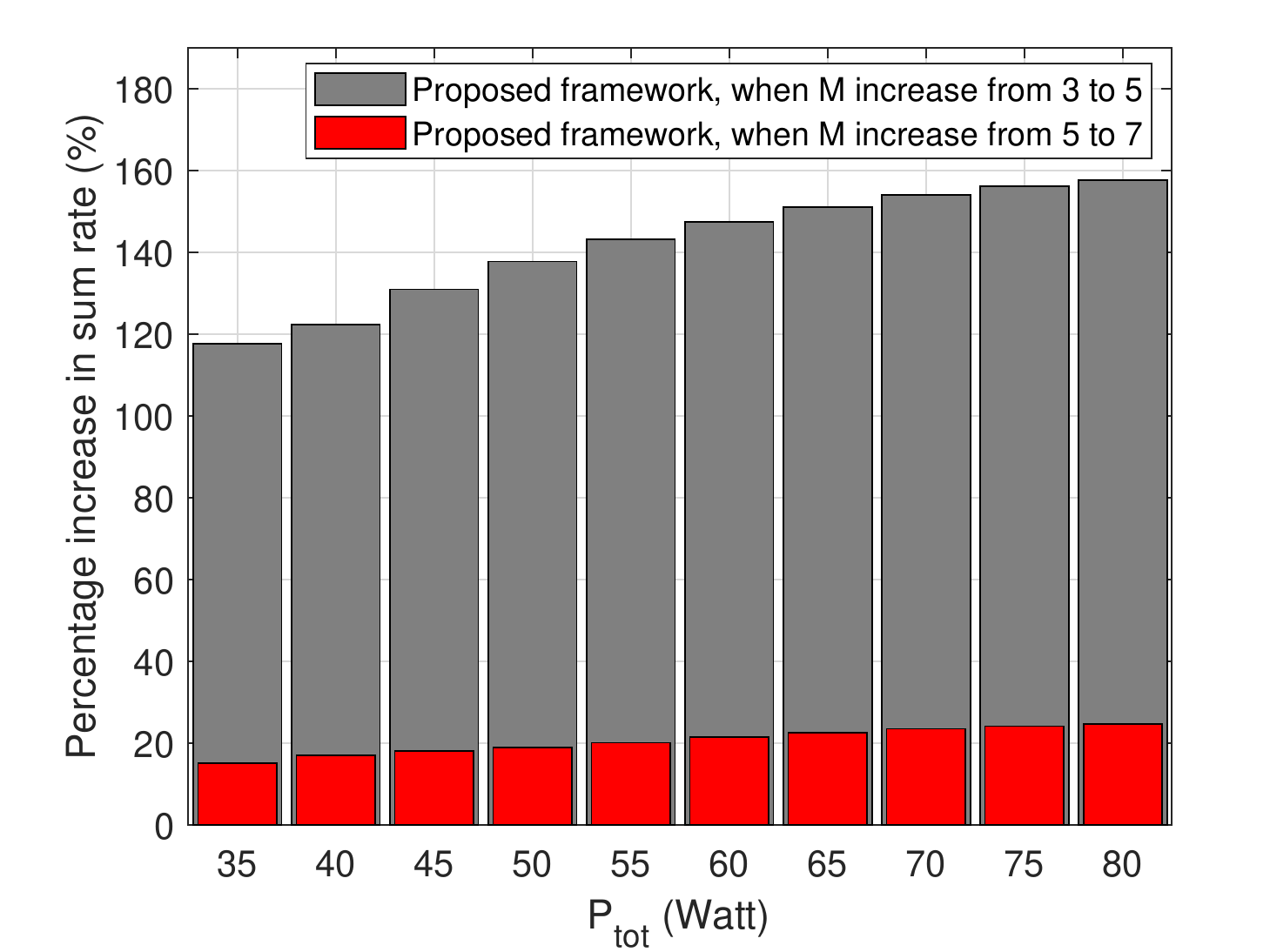}
\caption{Percentage gain in sum rate by the proposed framework when changes the number of beams at LEO SatCom.}
\label{rf33}
\end{figure}

The impact of number of beams $M$ on the sum rate of the system using the \textit{proposed framework} is shown in Fig. \ref{rf4}. It can be seen from the figure that the \textit{proposed framework} with more beams provides better performance compared to those with less beams. It is because more $M$ accommodate more RSMA users which enhances the sum rate of SatCom network. Further, when $P_{tot}$ is increased after a certain point, there is no improvement in the system performance because the transmission power of each beam ($p_{m,k}$) becomes bounded by the interference threshold ($I_{th}$). However, the point where increasing $P_{tot}$ has no impact on the sum rate (convergence like behaviour) comes sooner for the systems with smaller value of $M$. Besides, Fig. \ref{rf33} shows that a significant impact in performance by \textit{proposed framework} is achieved when the value of $M$ is increased from 3 to 5. For example, the percentage gain in sum rate by adding just two beams is more than 100\%. However, when the value of $M$ is increased from 5 to 7, the percentage increase in the sum rate is far less. Further, the percentage increase in the sum rate by applying \textit{proposed framework} increase with the increasing $P_{tot}$. This is because the system with smaller $M$ shows convergence like behaviour (where increasing $P_{tot}$ little impact on the sum rate) for smaller values of $P_{tot}$. Therefore, it becomes more beneficial to have larger values of $P_{tot}$ when the LEO satellite has larger number of beams.    
%%%%%%%%%%%%%%%%
%%%%%%%%%%%%%

\section{Conclusion}
Cognitive radio and RSMA have the potential to provide massive connectivity in space-ground communication networks. This paper has proposed RSMA for cognitive radio GEO-LEO coexisting satellite networks. Specifically, a new solution for maximizing the spectral efficiency of the secondary LEO system has been provided. The proposed framework has simultaneously optimized the transmit power of all beams, RSMA based power allocation over each beam, and resource block user assignment subject to each user's minimum rate and interference temperature to GEO ground users. To handle the non-convex optimization problem, a successive convex approximation technique, KKT conditions, and greedy-based algorithm have been adopted to obtain the efficient solution. For fair comparison, we proposed two benchmark optimization frameworks. Numerical results demonstrate that the proposed optimization scheme significantly improves the system performance with reasonable complexity. The system proposed in this work can be extended in several ways. One of the possible extensions would be considering multiple GEO and LEO satellites such that GEO satellites act as the primary network and LEO satellites act as the secondary network. In such a scenario, high levels of interference from primary to secondary to primary and from primary to secondary would be more challenging to manage. The possible solution could be applying reinforcement learning to handle such complex problems, which might be very hard to solve through conventional techniques.
%One can also be noted that the proposed model has been considered a single antenna for communication. Another exciting research is investigating the proposed model under the case when the devices are equipped with more than one antenna.

\ifCLASSOPTIONcaptionsoff
  \newpage
\fi

%\bibliographystyle{IEEEtran}
\bibliographystyle{IEEEtran}% This is IEEEtran.bst file
\bibliography{Wali_EE}

\begin{IEEEbiography}
[{\includegraphics[width=1in,height=1.5in,clip,keepaspectratio]{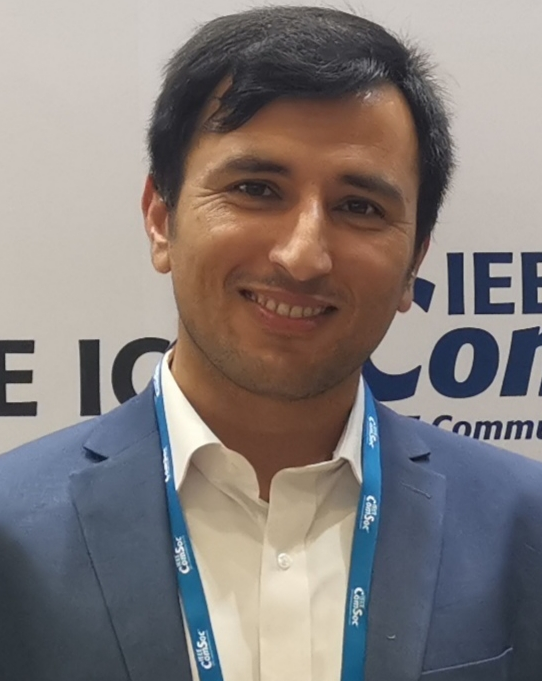}}]{Wali Ullah Khan} (Member, IEEE)
received the Master degree in Electrical Engineering from COMSATS University Islamabad, Pakistan, in 2017, and the Ph.D. degree in Information and Communication Engineering from Shandong University, Qingdao, China, in 2020. He is currently working with the Interdisciplinary Centre for Security, Reliability and Trust (SnT), University of Luxembourg, Luxembourg. He has authored/coauthored more than 70 publications, including international journals, peer-reviewed conferences, and book chapters. His research interests include convex/nonconvex optimizations, non-orthogonal multiple access, reflecting intelligent surfaces, ambient backscatter communications, Internet of things, intelligent transportation systems, satellite communications, unmanned aerial vehicles, physical layer security, and applications of machine learning.
\end{IEEEbiography}

\begin{IEEEbiography}
[{\includegraphics[width=1in,height=1.5in,clip,keepaspectratio]{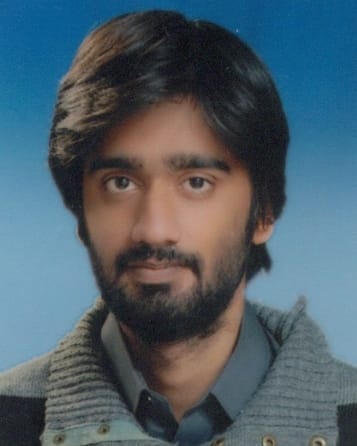}}]{Zain Ali} received his PhD degrees in Electrical Engineering from CIIT, Islamabad, Pakistan, in 2020. Currently, he is working as a Postdoc researcher at the University od California, Santa Cruz, CA, USA. His research interests include cognitive radio networks, energy harvesting, multi-hop relay networks, orthogonal frequency division multiple access (OFDMA), non-orthogonal multiple access (NOMA), satellite communication, machine learning and engineering optimization. He was awarded HEC's indigenous talent scholarship for MS and Ph.D. studies.
\end{IEEEbiography}

\begin{IEEEbiography}
[{\includegraphics[width=1in,height=1.5in,clip,keepaspectratio]{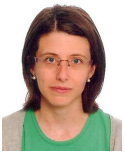}}]{Eva Lagunas} (Senior Member, IEEE)
received the M.Sc. and Ph.D. degrees in telecommunications engineering from the Polytechnic University of Catalonia (UPC), Barcelona, Spain, in 2010 and 2014, respectively. From 2009 to 2013, she was a Research
Assistant with the Department of Signal Theory and
Communications, UPC. In 2009 she was a Guest Research Assistant with the Department of Information
Engineering, University of Pisa, Pisa, Italy. From
November 2011 to May 2012, she held a Visiting
Research appointment with the Center for Advanced
Communications, Villanova University, PA, USA. In 2014, she joined the
Interdisciplinary Centre for Security, Reliability and Trust (SnT), University
of Luxembourg, Luxembourg, where she currently holds a Research Scientist
position. Her research interests include radio resource management and general
wireless networks optimization.
\end{IEEEbiography}

\begin{IEEEbiography}
[{\includegraphics[width=1in,height=1.5in,clip,keepaspectratio]{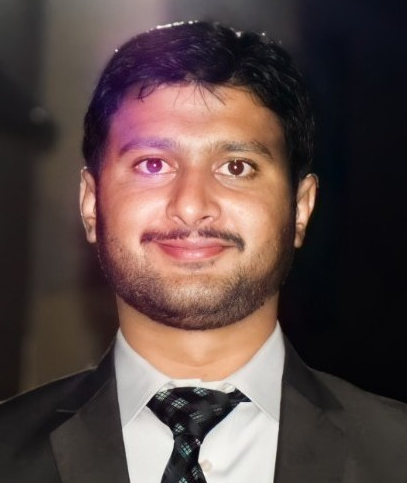}}]{Asad Mahmood} (Student Member, IEEE)
received his Master degrees in Electrical Engineering from the Department of Electrical \& Computer Engineering, COMSATS University Islamabad, Wah Campus, Pakistan. He is currently pursuing the Ph.D. degree with the Interdisciplinary Centre for Security, Reliability, and Trust (SnT), University of Luxembourg, where he is also a Doctoral Researcher. His research interest include resource allocation in UAV enabled wireless communication networks, Mobile Edge Computing, Machine learning, Evolutionary Algorithm, non-convex optimization and approximation algorithms for mixed integer programming in communication systems. 
\end{IEEEbiography}

\begin{IEEEbiography}
[{\includegraphics[width=1in,height=1.5in,clip,keepaspectratio]{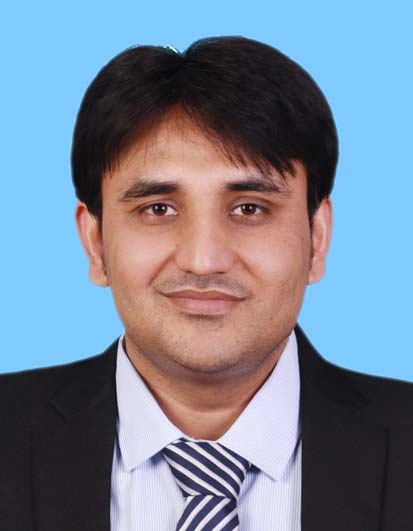}}]{Muhammad Asif} was born in Rahim Yar Khan, Bahawalpur Division, Pakistan, in 1990. He received the Bachelor of Science (B.Sc) degree in Telecommunication Engineering from The Islamia University of Bahawalpur (IUB), Punjab, Pakistan, in 2013, and Master degree in Communication and Information Systems from Northwestern Polytechnical University (NWPU), Xian, Shaanxi, China, in 2015. He also received Ph.D. degree in Information and Communication Engineering from University of Science and Technology of China (USTC), Hefei, Anhui, China in 2019. Currently, Dr. Asif is working as a post-doctoral researcher at the Department of Electronics and Information Engineering in Shenzhen University, Shenzhen, Guangdong, China. He has authored/co-authored more than 25 journal and conference papers. His research interests include Wireless Communication, Channel Coding, Coded-Cooperative Communication, Optimization and Resource Allocation, Backscatter-Enabled Wireless Communication, IRS-Assisted Next-generation IoT Networks.
\end{IEEEbiography}

\begin{IEEEbiography}
[{\includegraphics[width=1in,height=1.5in,clip,keepaspectratio]{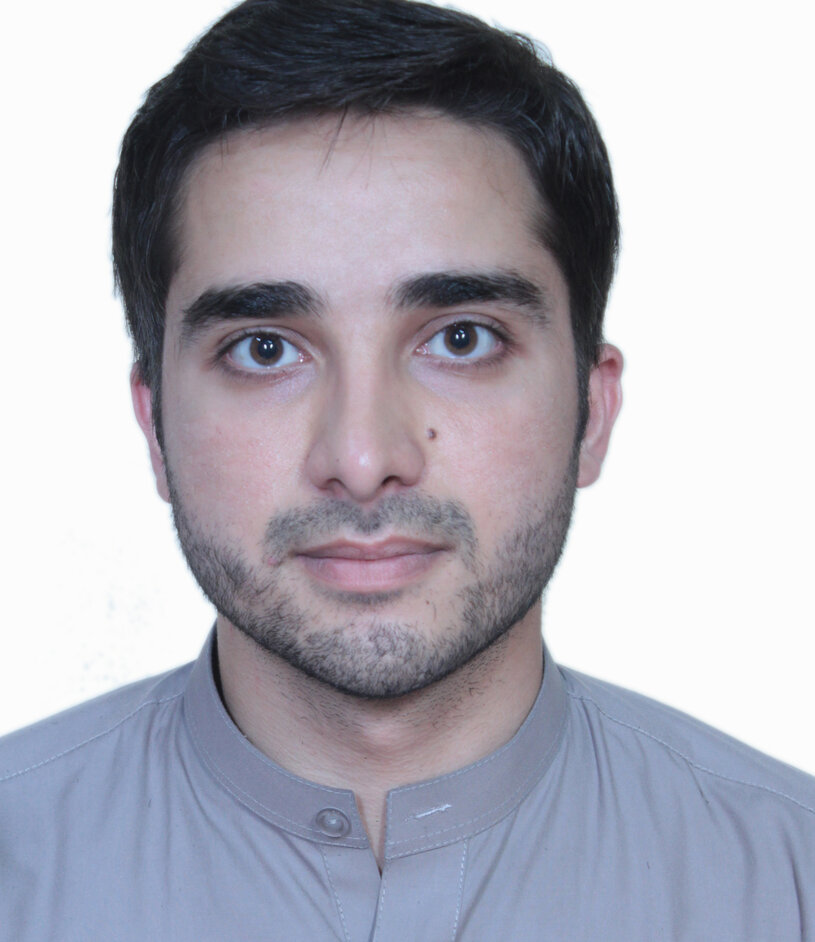}}]{Asim Ihsan} received the master’s degree in information and communication engineering from Xian Jiaotong University, Xi'an, China, and the Ph.D. degree in information and communication engineering from Shanghai JiaoTong University,
Shanghai,China. He is currently working as a Post-Doctoral Research Officer with the School of Computer Science and Electronic Engineering, Bangor University,U.K. He is also a Global Talent Visa Holder of U.K. His research interests include energy-efficient resource allocations for beyond 5G wireless communication technologies through convex/nonconvex optimizations and machine learning.
\end{IEEEbiography}

\begin{IEEEbiography}
[{\includegraphics[width=1in,height=1.5in,clip,keepaspectratio]{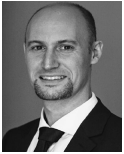}}]{Symeon Chatzinotas} (Fellow, IEEE)
received the M.Eng. degree in telecommunications from the Aristotle University of Thessaloniki, Thessaloniki, Greece, in 2003, and the M.Sc. and Ph.D. degrees in electronic engineering from the University of Surrey, Guildford, U.K., in 2006 and 2009,
respectively. He is currently a Full Professor or Chief
Scientist I and the Co-Head of the SIGCOM Research Group, Interdisciplinary Centre for Security, Reliability and Trust, University of Luxembourg. In the past, he was a Visiting Professor with the University of
Parma, Parma, Italy, and he was involved in numerous Research and Development projects for the National Center for Scientific Research Demokritos, the Center of Research and Technology Hellas and the Center of Communication
Systems Research, University of Surrey. He has coauthored more than 400
technical papers in refereed international journals, conferences and scientific
books. He was the co-recipient of the 2014 IEEE Distinguished Contributions
to Satellite Communications Award, the CROWNCOM 2015 Best Paper Award,
and the 2018 EURASIP JWCN Best Paper Award. He is currently in the
Editorial Board of the IEEE OPEN JOURNAL OF VEHICULAR TECHNOLOGY and
the International Journal of Satellite Communications and Networking.
\end{IEEEbiography}

\begin{IEEEbiography}
[{\includegraphics[width=1in,height=1.5in,clip,keepaspectratio]{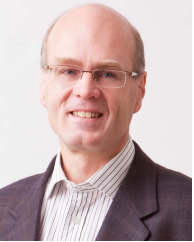}}]{Bj\"orn Ottersten} (Fellow, IEEE)
was born
in Stockholm, Sweden, in 1961. He received the
M.S. degree in electrical engineering and applied
physics from Linkoping University, Link ¨ oping, Swe- ¨
den, in 1986, and the Ph.D. degree in electrical
engineering from Stanford University, Stanford, CA,
USA, in 1990. He has held Research positions with
the Department of Electrical Engineering, Linkoping ¨
University, Linkoping, Sweden, the Information Sys- ¨
tems Laboratory, Stanford University, the Katholieke
Universiteit Leuven, Leuven, Belgium, and the University of Luxembourg, Esch-sur-Alzette, Luxembourg. From 1996 to 1997,
he was the Director of Research with ArrayComm, Inc., a Start-Up in San
Jose, CA, USA, based on his patented technology. In 1991, he was appointed
Professor of signal processing with the Royal Institute of Technology (KTH),
Stockholm, Sweden. He has been the Head of the Department for Signals,
Sensors, and Systems, KTH, and the Dean of the School of Electrical
Engineering, KTH. He is currently the Director for the Interdisciplinary Centre
for Security, Reliability and Trust, the University of Luxembourg. He was the
recipient of the IEEE Signal Processing Society Technical Achievement Award
and been twice awarded the European Research Council Advanced Research
Grant. He has coauthored journal papers which was the recipient of the IEEE
Signal Processing Society Best Paper Award in 1993, 2001, 2006, 2013, and
2019, and eight IEEE conference papers best paper awards. He has been a
Board Member of IEEE Signal Processing Society and the Swedish Research
Council and currently serves on the Boards of EURASIP and the Swedish
Foundation for Strategic Research. He was an Associate Editor for the IEEE
TRANSACTIONS ON SIGNAL PROCESSING and the Editorial Board of
the IEEE Signal Processing Magazine. He is currently a Member of the
Editorial Boards of the IEEE OPEN JOURNAL OF SIGNAL PROCESSING,
EURASIP Signal Processing Journal, EURASIP Journal of Advanced Signal
Processing and Foundations and Trends of Signal Processing. He is a Fellow
of EURASIP.
\end{IEEEbiography}

\begin{IEEEbiography}
[{\includegraphics[width=1in,height=1.5in,clip,keepaspectratio]{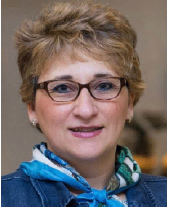}}]{OCTAVIA A. DOBRE} (Fellow, IEEE) received the Dipl.Ing. and Ph.D. degrees from the Polytechnic Institute of Bucharest, Bucharest, Romania, in 1991 and 2000, respectively. From 2002 to 2005, she was with the New Jersey Institute of Technology, Newark, NJ, USA. In 2005, she joined Memorial University, St. John’s, NL, Canada, where she is currently a Professor and the Research Chair. She was a Visiting Professor with the Massachusetts Institute of Technology, Cambridge, MA, USA, and the Universit de Bretagne Occidentale, Brest, France. She has authored coauthored more than 400 refereed papers in her research areas, which include wireless communication and networking technologies, and optical and underwater communications. She was the recipient of the Best Paper Awards at various conferences, including the IEEE ICC, the IEEE Globecom, the IEEE WCNC, and the
IEEE PIMRC. She is the Director of Journals and the Editor-in-Chief of IEEE OPEN JOURNAL OF THE COMMUNICATIONS SOCIETY. She was the Editor-in-Chief of IEEE COMMUNICATIONS LETTERS, a Senior Editor,
an Editor, and a Guest Editor of various prestigious journals and magazines. She was also the General Chair, the Technical Program Co-Chair, the Tutorial Co-Chair, and the Technical Co-Chair of symposia at numerous
conferences. She was a Fulbright Scholar, a Royal Society Scholar, and a Distinguished Lecturer of the IEEE Communications Society. She is an Elected Member of the European Academy of Sciences and Arts and a
Fellow of the Engineering Institute of Canada and the Canadian Academy of Engineering.
\end{IEEEbiography}

\end{document}